\documentclass{article}

\usepackage{arxiv}

\usepackage[utf8]{inputenc} 
\usepackage[T1]{fontenc}    
\usepackage{hyperref}       
\usepackage{subfig}
\usepackage{listings}
\usepackage{multirow}
\usepackage{xcolor}
\usepackage{amsmath}
\usepackage{makecell}
\usepackage{url}            
\usepackage{booktabs}       
\usepackage{amsfonts}       
\usepackage{nicefrac}       
\usepackage{microtype}      
\usepackage{array}
\newcolumntype{C}[1]{>{\centering\arraybackslash}m{#1}}
\newcolumntype{L}[1]{>{\raggedright\arraybackslash}m{#1}}
\usepackage{lipsum}
\usepackage{graphicx}
\lstdefinestyle{myCode}{
    backgroundcolor=\color{white},
    basicstyle=\ttfamily\scriptsize,
    frame=single,
    rulecolor=\color{black},
    frameround=tttt,
    breaklines=true,
    columns=fullflexible,
    escapeinside={(*@}{@*)} 
}
\graphicspath{ {./images/} }

\title{TIT: A Tree-Structured Instruction Tuning Approach for LLM-Based Code Translation}

\author{
 He Jiang \\
  School of Software\\
  Dalian University of Technology\\
  Dalian 116024, China \\
  \texttt{jianghe@dlut.edu.cn} \\
   \And
 Yufu Wang \\
  School of Software\\
  Dalian University of Technology\\
  Dalian 116024, China \\
  \texttt{wyf@mail.dlut.edu.cn} \\
  \And
 Hao Lin \\
  School of Software\\
  Dalian University of Technology\\
  Dalian 116024, China \\
  \texttt{dlutlh@mail.dlut.edu.cn} \\
  \And
 Peiyu Zou \\
  School of Computer Science and Artificial Intelligence\\
  Liaoning Normal University\\
  Dalian 116029, China \\
  \texttt{zoupeiyu@lnnu.cn} \\
  \And
 Zhide Zhou \\
  School of Software\\
  Dalian University of Technology\\
  Dalian 116024, China \\
  \texttt{cszide@gmail.com} \\
  \And
 Ang Jia \\
  School of Software\\
  Dalian University of Technology\\
  Dalian 116024, China \\
  \texttt{jiaang@dlut.edu.cn} \\
  \And
 Xiaochen Li \\
  School of Software\\
  Dalian University of Technology\\
  Dalian 116024, China \\
  \texttt{xiaochen.li@dlut.edu.cn} \\
  \And
 Zhilei Ren \\
  School of Software\\
  Dalian University of Technology\\
  Dalian 116024, China \\
  \texttt{zren@dlut.edu.cn} \\
}

\begin{document}
\maketitle
\begin{abstract}
Large Language Models (LLMs) have shown strong performance in automated source-to-target code translation through pretraining on extensive code corpora. However, mainstream LLM-based code translation methods suffer from two critical limitations. First, they are highly sensitive to language-specific features, which often introduce source-language syntax or lexicon into the output, leading to syntactic confusion. Second, they lack fine-grained semantic alignment due to an over-reliance on function-level parallel datasets, resulting in semantic misalignment between the translated code and the original source. To overcome these limitations, we propose TIT, a Tree-structured Instruction Tuning paradigm for LLM-based code translation. Specifically, TIT consists of three modules. First, to mitigate syntactic confusion, the syntactic information representation module integrates language-agnostic syntactic features via structured parsing. Then, to generate high-quality fine-grained parallel data, the fine-grained parallel dataset augmentation module aligns nodes with code segments through statement-level segmentation and contrastive matching. Finally, we leverage the dual-stage tree instruction tuning module to alleviate the contextual processing burden on the LLM caused by the introduction of syntactic information. The first stage employs syntax-aware fine-tuning to enable the LLM to autonomously comprehend structured syntactic information, while the second stage utilizes code generation fine-tuning to guide the model in generating accurate target code based on function-level syntactic dependencies. The experimental results demonstrate that the proposed method significantly outperforms existing approaches in multiple LLMs, achieving a success rate 1.22×-1.75× higher in code translation while markedly reducing syntactic confusion.
\end{abstract}

\section{Introduction}
\label{1}
Code translation, which involves converting source code from one programming language to another, has become a core requirement for cross-language software migration and legacy code reuse \cite{r1}. In the industrial sector, many legacy systems still rely on outdated programming language platforms, which face significant maintenance challenges and technical debt. To enhance system maintainability and adaptability, enterprises urgently need to migrate these legacy codes to modern mainstream platforms. Furthermore, with the rapid growth of multi-platform software applications, efficient cross-platform code translation has become a crucial demand in software development. Many industries spend hundreds of millions of dollars each year migrating code written in legacy programming languages (e.g., FORTRAN and COBOL) to modern languages like Java and C++ \cite{r7}, \cite{r8}. Traditional methods often rely on manual translation, hand-crafted rules, and fixed templates, which are not only costly but also struggle to handle complex syntax and language-specific features\cite{r2}. To address these challenges, researchers have started to develop automated translation solutions with low implementation costs and high accuracy to improve translation efficiency and reduce reliance on manual intervention\cite{transcoder}, \cite{tree-tree}, \cite{dobf}.

With significant advancements in deep learning, particularly the emergence of the Transformer architecture \cite{transformer}, pre-trained LLMs have become the primary approach for code translation tasks, demonstrating exceptional performance. Typical end-to-end approaches \cite{codebert,codet5} directly feed source code into LLMs and enhance training efficiency by incorporating machine learning components, such as attention mechanisms or adversarial training. To reduce the dependence on large-scale parallel datasets, recent approaches have leveraged Intermediate Representations (IR) for code translation. For example, Chen et al. \cite{tree-tree} separated syntax learning from syntax-to-code alignment by using abstract syntax trees (ASTs) as IR, simplifying the translation process. Niu et al. \cite{spt-code} further integrated linearized AST representations into LLM pretraining objectives, alleviating the scarcity of parallel datasets. These methods enhance LLMs’ understanding of source code and improve translation robustness by providing more detailed information about syntax and semantics, enabling high-quality translation even with relatively limited parallel datasets.

Despite these advances, achieving fine-grained translation that is both syntactically accurate and consistent with target language standards remains challenging. We identify two critical limitations:

\begin{itemize}
    \item{Sensitivity to language-specific features: LLMs are often sensitive to interference from source-language syntax or vocabulary during inference, which may produce outputs that confuse the syntactic rules of both source and target languages. We define this phenomenon as the syntactic confusion problem. Empirical studies \cite{lost-in-translation} show that LLMs frequently fail to distinguish differences between the syntactic and semantic systems of source and target languages. Consequently, translated outputs may inappropriately retain source-language elements—such as package names, function identifiers, or structural patterns—or even continue portions of the input code. Such syntactic confusion undermines the correctness and standard compliance of generated target code, highlighting the need to preserve important semantic and structural information while minimizing source-language interference.}
    \item{Lack of fine-grained semantic alignment: The accuracy of code translation also depends on the alignment granularity of parallel datasets \cite{xlcost}. Existing researches \cite{avatar}, \cite{CodeTransOcean}, \cite{G-TransEval} typically constructs function-level datasets from programming competition platforms, using solutions in different languages for the same problem as parallel data. Although these implementations perform the same algorithmic functions, their statement sequences often differ significantly, causing semantic misalignment. This prevents pre-trained models from learning fine-grained cross-lingual representations. Therefore, statement-level parallel datasets are essential for LLMs to capture precise semantic alignments and achieve high-quality translation.}
\end{itemize}

To address these limitations, we propose TIT, a tree-structured instruction tuning method specifically designed for LLMs involved in code translation. For the first limitation, TIT introduces a syntactic information representation module that optimizes language-specific nodes in the source code's AST through pruning and information replacement. This module preserves key syntactic information in the input to the LLM while preventing the model from continuing the source code during inference, thereby mitigating syntactic confusion. To address the second limitation, TIT proposes a fine-grained parallel dataset augmentation module. By segmenting function-level parallel data into statement-level code snippets based on AST nodes and applying a node-to-snippet matching model trained on limited statement-level parallel dataset, this module filters high-confidence aligned samples to augment fine-grained parallel datasets in both scale and quality. However, introducing syntactic information representations concurrently induces excessively long contextual inputs. This impairs LLMs' reasoning capabilities and diminishes training efficacy. To address this issue, TIT designs a dual-stage tree instruction tuning module. In the first instruction fine-tuning stage, TIT guides LLMs to align syntactic information representation nodes with corresponding target-language code snippets through syntax-aware fine-tuning and fine-grained parallel datasets during self-supervised learning. This process enables autonomous discovery of associations between syntactic structures and target code. In the subsequent second stage, TIT performs function-level fine-tuning on the syntax-aware LLM from the previous stage via code generation tasks. This optimizes the LLM's capacity to integrate contextual dependencies of code snippets and improves its capacity to generate executable target code from syntactic information representations.

To evaluate the effectiveness and generalizability of TIT, we applied it across code-LLMs with various architectures and parameter scales, including StarCoder2-15B \cite{starcoder2}, StarCoder2-7B and CodeQwen1.5-7B \cite{codeqwen}, conducting comprehensive evaluations on the HumanEval-X \cite{humaneval-x} benchmark. Our results demonstrate that TIT achieves relative successful translation rates of 1.75×, 1.45×, and 1.22× from the respective base models, while reducing syntactic confusion rates by 72.73\%, 17.65\%, and 80.00\%. 

We further compared TIT-integrated LLMs with state-of-the-art specialized code translation methods and general-purpose LLMs. The results indicate that TIT-integrated LLMs consistently outperform specialized code translation methods across evaluation metrics. When compared to general-purpose models with substantially larger parameter scales, TIT achieves competitive performance and even surpasses certain counterparts. For instance, TIT-integrated StarCoder2-15B reaches a 68.29\% translation success rate – comparable to GPT-4 Turbo \cite{gpt4} (70.12\%) and DeepSeek-V3 \cite{deepseekv3} (71.95\%), while exceeding DeepSeek-V2 \cite{deepseekv2} (59.15\%).

To summarize, TIT made the following contributions.

\begin{enumerate}
\item{To the best of our knowledge, we are the first to leverage language-agnostic syntactic information representation to address the issue of syntactic confusion in LLM-based code translation. This approach significantly improves translation accuracy and enhances the utilization of syntactic information.}
\item{We propose a novel tree-structured instruction tuning paradigm named TIT, which introduces a syntactic information representation module, a fine-grained parallel dataset augmentation module, and a dual-stage tree instruction tuning module. These modules enable LLMs to comprehend and utilize syntactic and semantic information at a fine-grained level, thereby improving code translation accuracy while effectively mitigating the impact of excessively long inputs on the model’s inference ability.}
\item{We conducted extensive experiments using multiple large language models and benchmark code translation datasets to evaluate the effectiveness of the proposed approach. Results demonstrate that TIT achieves superior performance in improving translation accuracy and reducing syntactic confusion, while maintaining computational efficiency.}
\end{enumerate}

\begin{figure*}[!t]
\centering
\subfloat[]{\includegraphics[width=2.5in]{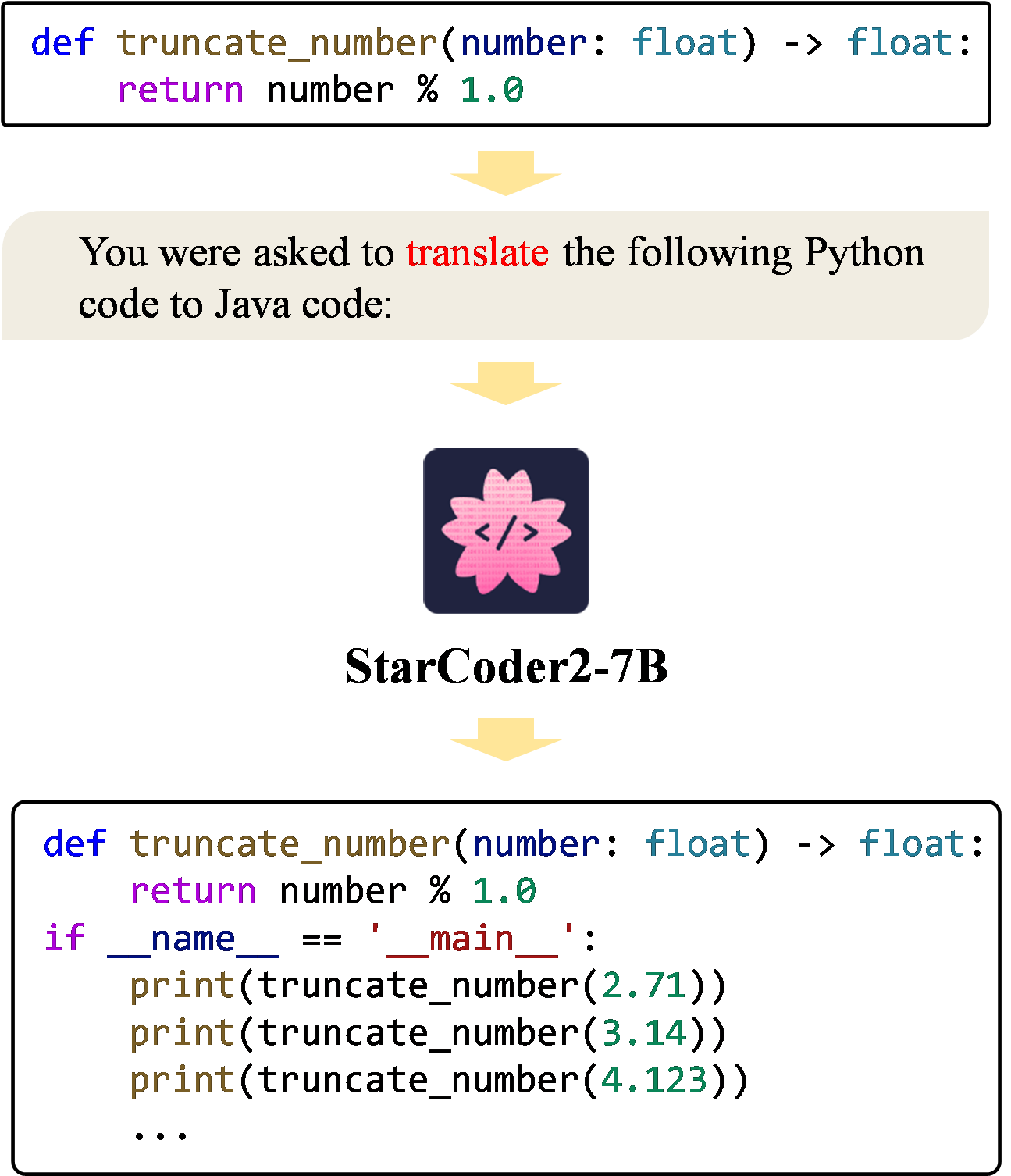}%
\label{fig_sum_1_first_case}}
\hfil
\subfloat[]{\includegraphics[width=2.5in]{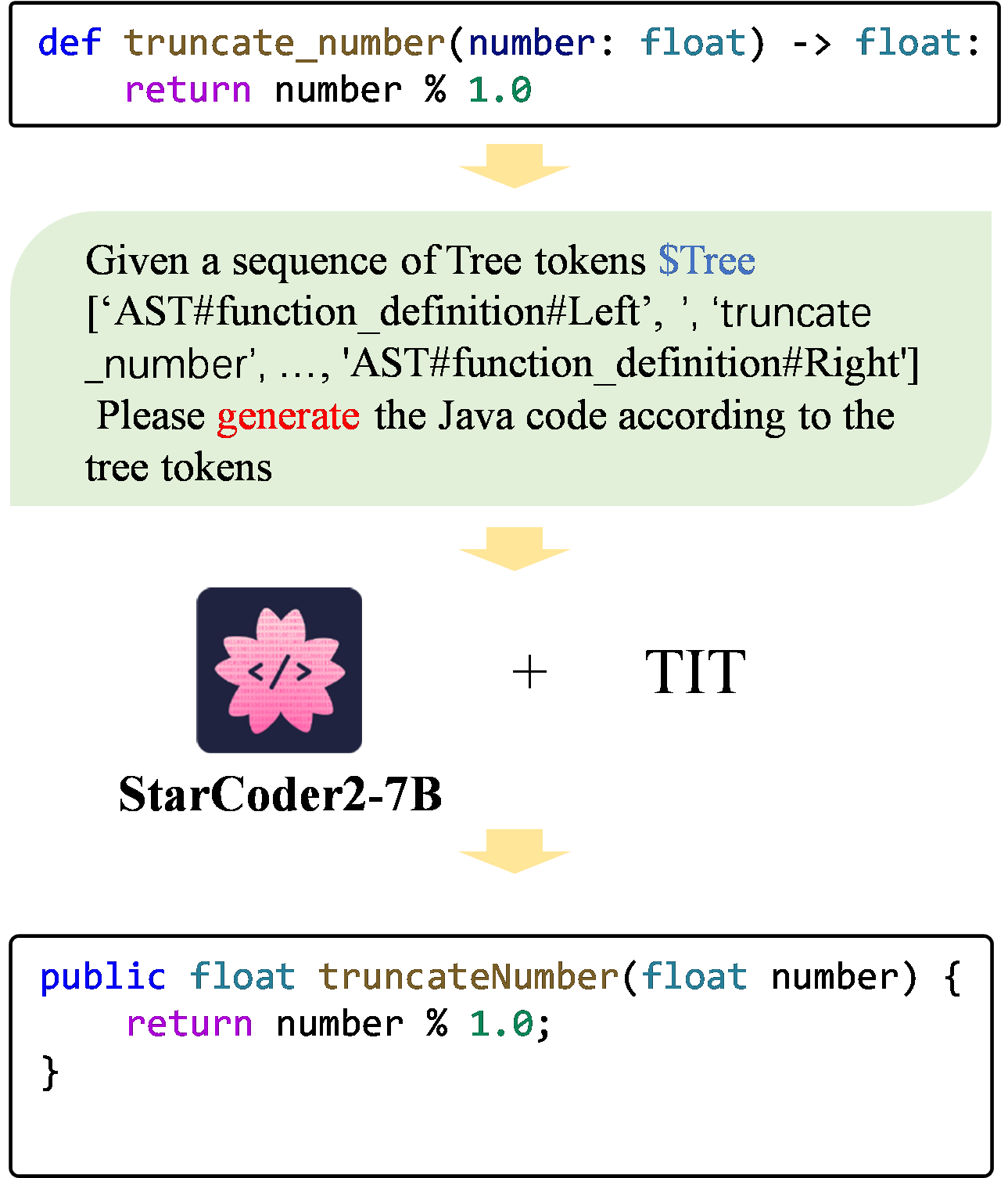}%
\label{fig_sum_1_second_case}}
\caption{A motivation example demonstrates the TIT's potential to mitigate the syntactic confusion. (a) Evaluation on StarCoder2-7B without TIT. (b) Evaluation on StarCoder2-7B with TIT.}
\label{fig_sum_1}
\end{figure*}

\section{Motivation}
\subsection{Syntactic Confusion}

As discussed in Section \ref{1}, syntactic confusion occurs when LLMs either continue source code content in translation outputs or include source-language syntactic tokens within target code. Fig. \ref{fig_sum_1}\subref{fig_sum_1_first_case} provides a motivation example demonstrating syntactic confusion in LLM-based code translation. In this case, the translation task involves converting the Python source code of the function \texttt{truncate\_number} into Java. This function accepts a float-type parameter called \texttt{number}, performs a modulo operation on the input, and returns the modulo result as a float-type value. The evaluated model StarCoder2-7B failed to recognize the translation objective. Instead, it extended the function by generating a main method with test cases.

\begin{figure}[!t]
\centering
\includegraphics[width=5in]{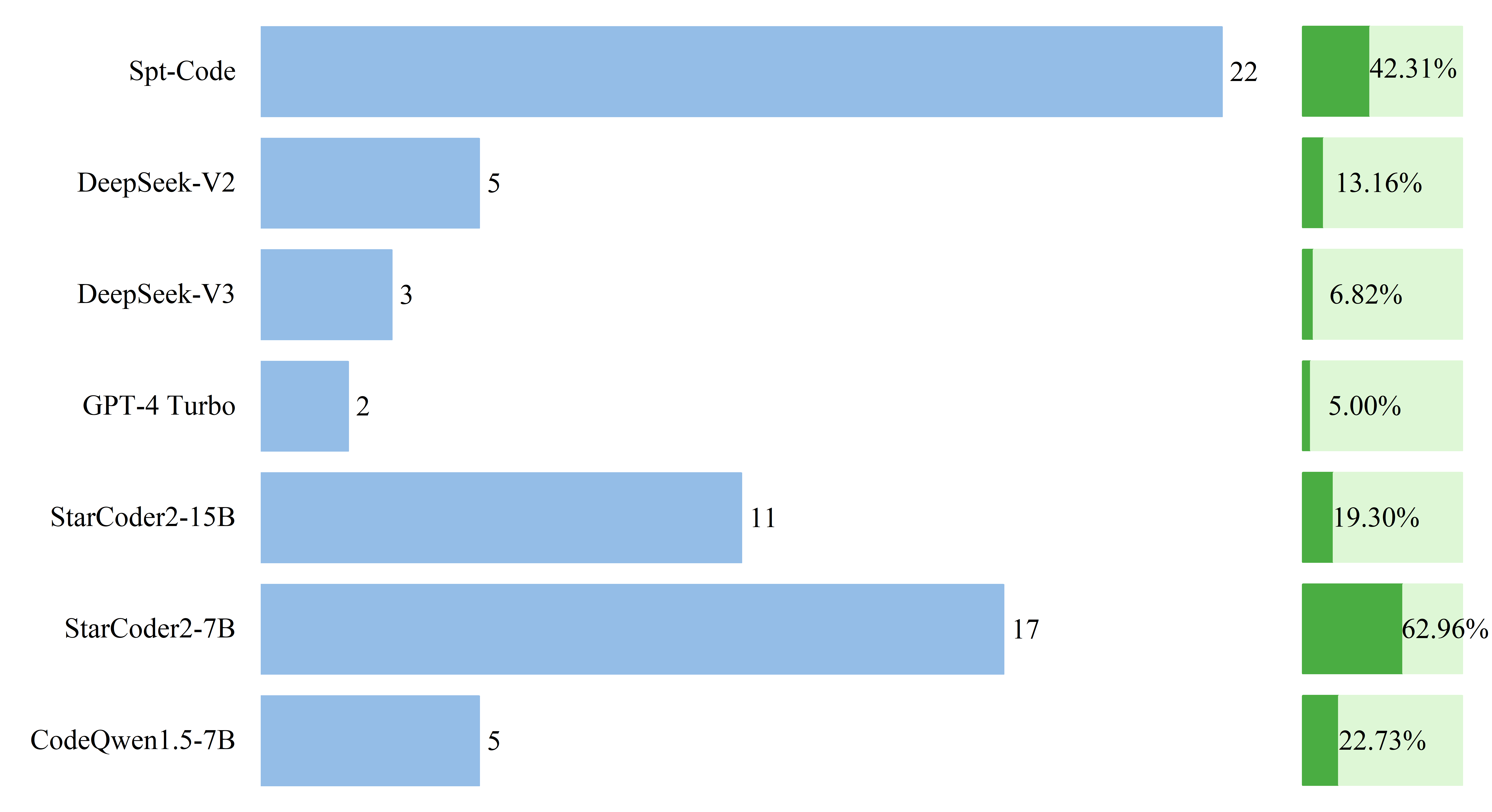}
\caption{Syntactic confusion across methods on the HumanEval-X Dataset.}
\label{fig_2}
\end{figure}

We evaluated state-of-the-art LLMs and code translation methods on the code translation task, measuring the proportion of compilation errors caused by syntactic confusion, with the results shown in Fig. \ref{fig_2}. Experimental details are provided in Section \ref{4}. All evaluated LLMs exhibited syntactic confusion at rates between 4.08\% and 32.69\%, varying by model architecture and parameter scale. Notably, TransCoder showed 0\% syntactic confusion due to its exclusive use of monolingual training data, which intrinsically prevents this cross-lingual continuation phenomenon.

Our analysis indicates that syntactic confusion stems from misleading source-language tokens in LLM inputs. When processing user instructions, LLMs conflate ``source code generation task'' with ``source-to-target translation task'', resulting in rewritten or extended source code contents. In Fig. \ref{fig_sum_1}\subref{fig_sum_1_second_case}, we use linearized AST representations instead of source code to provide enhanced syntactic information. By pruning and replacing language-specific nodes, the evaluated model StarCoder2-7B effectively performs the code translation task when employed with TIT.

Enlightened by the above finding in the analysis, we propose using a language-agnostic AST as an intermediate representation to provide critical information for code translation tasks. This approach is guided by three primary considerations. First, the AST is one of the most common intermediate results in the compilation process and aligns more closely with syntactic structures, helping to reduce differences in representation between programming languages. Second, LLMs have already acquired extensive syntactic knowledge through pretraining on large-scale code corpora. Compared to relying on raw code structures, which can include distracting signals, LLMs can more effectively leverage such intermediate representations to generate target code. Finally, from a statistical perspective, emphasizing key information from intermediate representations increases the probability that LLMs will correctly associate tokens with the target language during training. Based on these insights, we introduce a syntactic information representation module that leverages processed language-agnostic linearized ASTs as an intermediate representation. This prevents language-specific tokens from interfering with the LLMs during code translation, thereby mitigating syntactic confusion. The details of this module will be presented in the next section.

\begin{figure}[!t]
\centering
\includegraphics[width=5in]{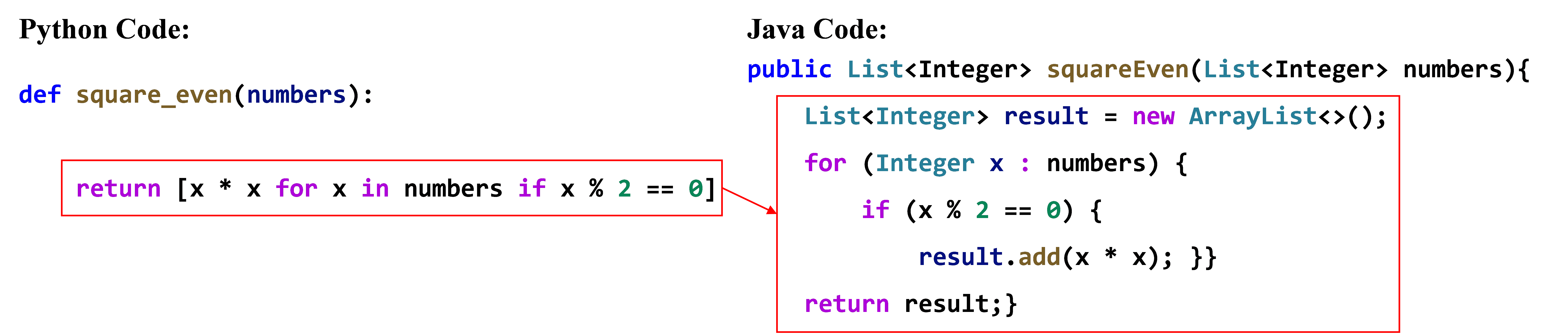}
\caption{An example of line-efficiency non-equivalence in expression density within function-level parallel datasets.}
\label{fig_4}
\end{figure}

\subsection{Semantic Misalignment}
Another critical concern is semantic misalignment, where translated code exhibits functional divergence from the source logic despite syntactic validity. The following example demonstrates boundary management issues in LLM-based translation of reverse traversal code. Python's \texttt{range(start, stop, step)} function inherently handles index boundary adjustments. When directly translated by LLMs into Java for loops, this approach may introduce potential \texttt{ArrayIndexOutOfBoundsException} errors.

\begin{lstlisting}[style=myCode]
for i in range(len(arr)-1, -1, -1): # Original Python code
for (int i = arr.length; i >= 0; i--) # Incorrect Java code
\end{lstlisting}

The main issue stems from the lack of fine-grained parallel datasets. Current studies typically construct function-level datasets from programming competition platforms, where solutions in different languages for the same problem are regarded as function-level parallel datasets. This approach suffers critical limitations: despite functionally equivalent, implementations exhibit formal non-equivalence in structure, sequential non-equivalence in execution order, and line-efficiency non-equivalence in expression density. Fig. \ref{fig_4} provides an example of line-efficiency non-equivalence in expression density from the function-level parallel datasets. In this case, Python leverages its powerful list comprehension syntax to implement the \texttt{square\_even} function in a single line of code, while Java requires five lines. Therefore, pursuing statement-level parallelism in function-level parallel datasets is generally unattainable and often misguided. Furthermore, incorporating such datasets in LLM training impedes the model's ability to capture fine-grained semantic relationships, potentially inducing runtime errors.

In summary, fine-grained parallel data is essential for developing LLMs' fine-grained semantic comprehension. Presently, the only available statement-level dataset, XLCoST, provides merely 88,954 statement pairs from GeeksForGeeks\footnote{https://www.geeksforgeeks.org/}, a platform that hosts thousands of data structure and algorithm problems.

To mitigate semantic misalignment, we augment fine-grained parallel datasets at the statement level, enabling LLMs to more accurately capture syntax-semantic correspondences between source and target code in fine-grained translation tasks. Detailed procedures will be provided in the next section.

\begin{figure*}[!t]
\centering
\includegraphics[width=6in]{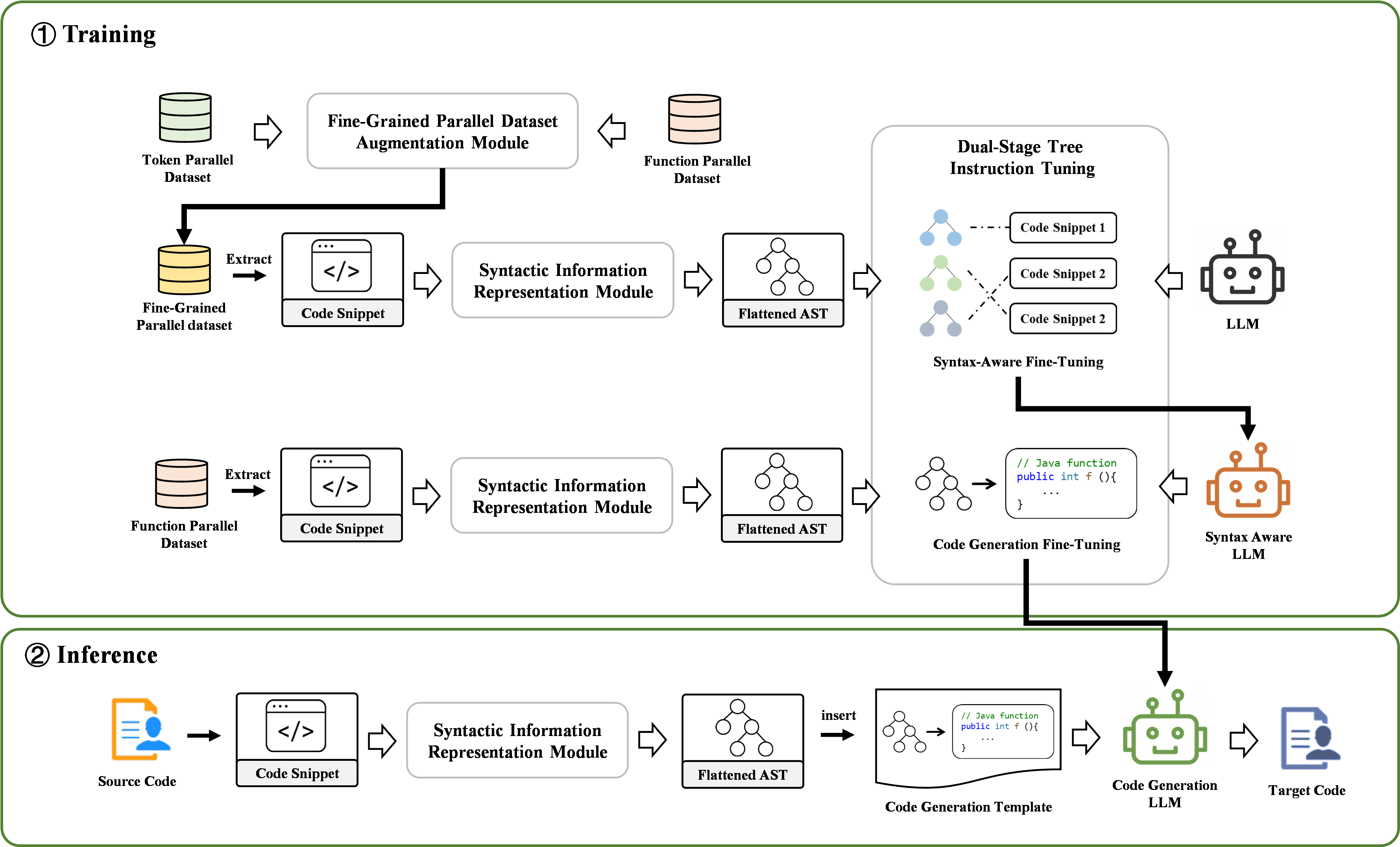}
\caption{Workflow of TIT.}
\label{fig_5}
\end{figure*}

\section{Method}
\subsection{Overview}
TIT consists of three main modules: a syntactic information representation module, a fine-grained parallel dataset augmentation module, and a two-level tree instruction tuning module. The syntactic information representation module generates a language-agnostic representation of syntactic information by parsing source code into a linearized AST and applying rule-based templates for language-specific node pruning and replacement. This module ensures both the sufficiency and purity of syntactic information. The fine-grained parallel dataset augmentation module trains a tree-to-code alignment matching model on a small-scale statement-level parallel dataset, segments function-level datasets, and filters high-confidence AST node-to-code snippet alignments, thereby enhancing the quality and scale of fine-grained parallel datasets. To address the challenge posed by long inputs containing rich syntactic information during LLM training and inference, TIT introduces the dual-stage tree instruction tuning module. In the first stage, TIT employs self-supervised syntax-aware fine-tuning to guide the LLM in accurately mapping tree-structured syntactic information representation nodes to code snippets using fine-grained parallel data. In the second stage, TIT incorporates code generation fine-tuning, encouraging the LLM to generate accurate target code based on complete syntactic information representations, while improving context-dependent integration and code synthesis efficiency. This progressive dual-stage fine-tuning strategy effectively balances the optimization of local syntactic alignment and global code generation.

Fig. \ref{fig_5} illustrates the interaction between TIT modules during training and inference. In the training phase, the function-level dataset is segmented into code snippets, which are then expanded into a fine-grained parallel dataset using the fine-grained parallel dataset augmentation module. The syntactic information representation module converts these datasets into structured representations, enabling syntax-aware fine-tuning of base models. The syntax-aware LLM is then fine-tuned for code generation on function-level datasets, producing a target code generation LLM capable of synthesizing executable code from syntactic information representations. In the inference phase, source code is processed by the syntactic information representation module to generate syntactic information representations, which are then embedded into code generation instruction templates and fed into the code generation LLM to produce target code.

\subsection{Syntactic Information Representation Module}
This module concentrates on generating syntax-agnostic descriptions of structural syntactic information for the source code, providing LLMs with precise and pure input. This representation serves as an IR for code translation. 

We obtain the AST corresponding to the source code by deploying a parser\footnote{https://tree-sitter.github.io/tree-sitter}, which expresses the syntactic information of the source code as a tree structure of "parameters (internal nodes) – data (leaf nodes)". Since the raw tree structure is incompatible with direct input to the LLM, we convert the AST into a linear form to improve its conciseness and processability. We adopt the linearization method proposed by \cite{unixcoder} to map the AST into a textual sequence. Specifically, given the root node of an AST, we recursively process its child nodes, representing internal and leaf nodes with the template ‘⟨parameters, left⟩, (data), ⟨parameters, right⟩’, while leaf nodes are directly denoted by their names.

\begin{table*}[h]
\centering
\caption{Formulation rules for the syntactic information representation.\label{tab:table1}}
\renewcommand{\arraystretch}{1.5}
\small
\begin{tabular}{C{4cm} L{5cm} C{2cm} C{3cm}}
\toprule
\makecell{TYPE} & \makecell{DEFINITION} & \makecell{OPERATION} & \makecell{EXAMPLE} \\
\midrule
Universal Nodes & With an identical structure and meaning across languages, it requires no special adaptation. & Unchange & \texttt{if\_statement} \\
Semi-Universal Nodes & Share common semantics but exhibit variant representations across languages. & Replace & \texttt{BlockStatement}→
\texttt{CompoundStatement}
 \\
Language-Specific Nodes & Only exist in specific languages and resist expression through generic node compositions. & Prune & \texttt{:} → ×
\\
\bottomrule
\end{tabular}
\end{table*}

To prevent interference from source-language-specific syntactic rules and tokens in the LLM input, we remove or generalize nodes containing such specialized information. Specifically, we prune language-specific syntactic tokens and expression parameters from ASTs. Concurrently, we normalize equivalent expressions across programming languages into a language-agnostic abstract representation. This process results in a purified syntactic information representation of source code, with formal construction rules defined in Table 1. The refined representation retains only expression parameters, operators, and identifier names essential for cross-lingual translation. 

\begin{figure}[!t]
\centering
\includegraphics[width=5in]{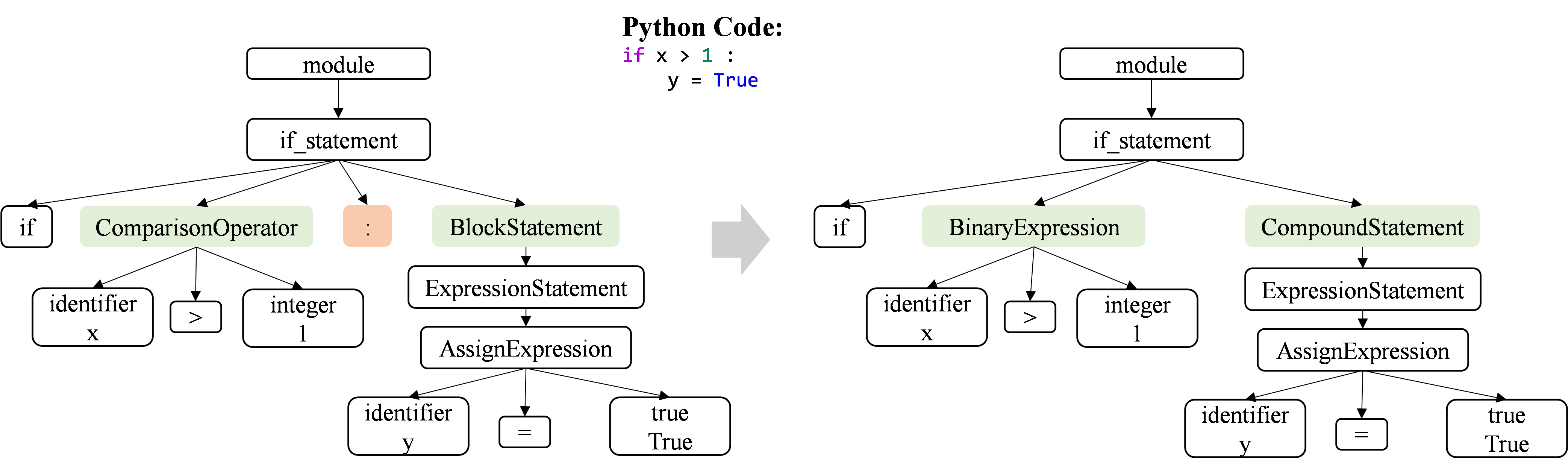}
\caption{An example of processing the AST node.}
\label{fig_6}
\end{figure}

We illustrate a specific example of processing the AST node in Fig. \ref{fig_6}. In this example, \texttt{if\_statement} exemplifies a universal node preserved in syntactic information representations due to cross-language consistency. Conversely, \texttt{BlockStatement} represents a semi-universal node. In Python's AST representation, \texttt{BlockStatement} indicates code blocks, which typically appear within conditional or loop statements to demarcate executable units. Unlike brace-delimited languages (e.g., C/Java), Python uses indentation instead of explicit \texttt{\{\}} tokens to define blocks. Therefore, we set \texttt{CompoundStatement} as the universal expression parameter for code blocks across languages' ASTs, replacing language-specific representations. As a language-specific node, the colon (\texttt{:}) in Python specifically signals the start of a code block. To maintain language-agnostic syntactic representations, we systematically prune this node. Instead, the retained \texttt{CompoundStatement} node explicitly captures the universal semantic role of code blocks, abstracting implementation-specific syntax while preserving structural functionality across all target languages.

\subsection{Fine-Grained Parallel Dataset Augmentation Module}
In this module, we focus on extracting finer-grained parallel datasets from existing function-level parallel datasets. This enables LLMs to effectively capture intricate syntactic details from tree-structured syntactic representations, while also reducing their computational burden during fine-tuning and reducing semantic alignment bias during inference.

We segment collected function-level parallel datasets into statement-level datasets based on node boundaries from tree-structured syntactic information representations. However, these segmented statements are not naturally functionally equivalent, primarily due to discrepancies in the original function-level parallel datasets caused by the order of implementation. Our method focuses on aligning the structured encodings of both syntactic representations and target code snippets to create fine-grained parallel pairs. To achieve this, we train a matching model using a small-scale statement-level parallel dataset proposed in \cite{xlcost}, combined with a multi-dimensional alignment approach for node-to-snippet encoding. Guided by this alignment approach, the matching model associates syntactic information representation nodes of source code with corresponding target code snippets, thereby augmenting the fine-grained parallel dataset. 

\textbf{Structured Encoding}. For each function-level sample in dataset \cite{xlcost}, the source code is decomposed into statement-level code snippets \(S = \{ {s_i}\} _{i = 1}^n\), while the target code is segmented into corresponding statement-level code snippets \(C = \{ {c_i}\} _{i = 1}^N\). Here, \({s_i}\) and \({c_i}\) maintain semantic and functional equivalence. To ensure compatibility with large language models, we employ a vanilla encoder \cite{transformer}, denoted as \({f_r}\), to generate structured encodings for both the nodes of the source syntactic information representation and the target code snippets. Element-wise L2 normalization is subsequently applied to each aligned node's textual encoding. The structured encodings of both syntactic representations \(\widehat P\) and target code \(\widehat X\) are represented as follows:


\begin{equation}
\label{eq1}
\begin{aligned}
P &= f_r(T),\; X = f_r(C), \\
\widehat P &= norm(P),\; \widehat X = norm(X)
\end{aligned}
\end{equation}

\textbf{Matching Model}. Inspired by prior works \cite{match_model}, we achieve structured encoding alignment between syntactic information representations and code snippets through three mechanisms: (i) Independent Encoding Alignment (equation (\ref{eq2})): Computes dot-product similarity between each tree node encoding \({\widehat P_i}\) and its corresponding code snippet encoding \({\widehat X_i}\), capturing direct node-to-snippet correspondences. (ii) Structural Encoding Alignment (equation (\ref{eq3})): Matches each tree node encoding \({\widehat P_i}\) with the mean encoding of \({\widehat X_i}\)’s local neighborhood, modeling syntactic contextual dependencies. (iii) Hybrid Encoding Alignment (equation (\ref{eq4})): Aligns each code snippet encoding \({\widehat X_i}\) with the mean encoding of parent/sibling nodes for \({\widehat P_i}\), incorporating bidirectional alignment considerations.

\begin{equation}
\label{eq2}
\begin{aligned}
g_1^{(1)}(\widehat P)\; &= \;{\rm{\{ }}\;{\widehat P_i}\;, 1 \le \;i\; \le \;N\} ,\\
g_1^{(2)}(\widehat X)\; &= \;{\rm{\{ }}\;{\widehat X_i}\;, 1 \le \;i\; \le \;N\}
\end{aligned}
\end{equation}

\begin{equation}
\label{eq3}
\begin{aligned}
g_2^{(1)}(\widehat P) & = {\rm\{} \widehat P_i \;, 1 \le i \le N {\rm\}},\\
g_2^{(2)}(\widehat X) & = {\rm\{} \frac{1}{|N_i|} \sum_{j \in N_i} \widehat X_j \;, 1 \le i \le N {\rm\}}
\end{aligned}
\end{equation}

\begin{equation}
\label{eq4}
\begin{aligned}
g_3^{(1)}(\widehat P)\; &= \;\{ \;{\widehat X_i}\;, 1 \le \;i\; \le \;N\} ,\\
g_3^{(2)}(\widehat X)\; &= \;\{ \;\frac{1}{|N_i|} \sum_{j \in N_i} {{{\widehat X}_j}} \;, 1 \le \;i\; \le \;N\}
\end{aligned}
\end{equation}

The matching model and loss function are constructed as follows:
\begin{equation}
\label{eq5}
{\Gamma _k}\; = \;(\;g_k^{(1)}(\widehat P)g_k^{(2)}{(\widehat X)^ \top })\; \cdot \;\exp ({\tau _{match}})
\end{equation}
\begin{equation}
\label{eq6}
L = \;\sum_k {\frac{1}{2}} {\lambda _k}(CE({\Gamma _k},\;y)\; + \;CE(\Gamma _k^ \top ,\;y))
\end{equation}

where \({\tau _{match}} \in \mathbb{R} \) denotes a trainable temperature parameter modulating similarity magnitudes, while \(g_k^{(1)}\) and \(g_k^{(2)}\) represent dimension-specific transformation functions formalized in equations (\ref{eq2}), (\ref{eq3}), (\ref{eq4}). \(N\) corresponds to the first dimension of \(\widehat P\) and \(\widehat X\), and \({\lambda _k} \in \mathbb{R}{^{\rm{ + }}}\)  serves as a balancing hyperparameter for multi-dimensional alignment mechanisms' contributions. The contrastive loss \(L\) for the similarity matrix \(\Gamma \) follows equation (\ref{eq6}), where \(CE\) denotes the cross-entropy function and \(y = {(0,\;1,\;...,\;N - 1)^ \top }\) defines the contrastive training label vector.

\textbf{Data Augmentation}. We collect existing function-level parallel datasets and remove duplicates. For each function-level sample, we decompose the source code into statement-level snippets \(S = \{ {s_i}\} _{i = 1}^M\) based on node boundaries of tree-structured syntactic representations, while segmenting the target code into candidate snippets \(C = \{ {c_q}\} _{q = 1}^N\). The syntactic information representation module generates corresponding tree-structured representations \(T = \{ {t_i}\} _{i = 1}^M\) for the source code. We compute a three-dimensional similarity matrix \({\Gamma ^{(k)}} = [\Gamma _k^{(1)},\;\Gamma _k^{(2)},\;\Gamma _k^{(3)}] \in \mathbb{R}{^{N \times {\rm{3}}}}\) between \({t_i}\) and candidate set \(C\) by integrating our encoding scheme with the matching model. We normalize \({\Gamma ^{(k)}}\) to obtain the \(k\)-th dimension similarity score \(s_{iq}^{(k)}\) between \({t_i}\) and \(q\)-th candidate element \({c_q}\), given by:
\begin{equation}
\label{eq7}
s_{iq}^{(k)}\; = \;\frac{{\exp (\Gamma _{iq}^{(k)}/{\tau _{norm}})}}{{\sum\limits_{m = 1}^N {\exp (\Gamma _{im}^{(k)}/{\tau _{norm}})} }}
\end{equation}
where \({\tau _{norm}}\) is the normalization temperature coefficient, we empirically set its value to 0.1 following \cite{adapt_parameter}. Candidate elements are selected only when achieving both the highest mean three-dimensional similarity score and exceeding the empirically optimal threshold of 0.85 following \cite{parameter_for_similarity}. These candidates are paired with corresponding nodes to form fine-grained parallel data pairs, while nodes without qualifying candidates are discarded. Finally, all valid pairs are consolidated into the refined fine-grained parallel dataset.

\subsection{Dual-Stage Tree Instruction Tuning Module}

\begin{figure*}[!t]
\centering
\subfloat[]{\includegraphics[width=3in]{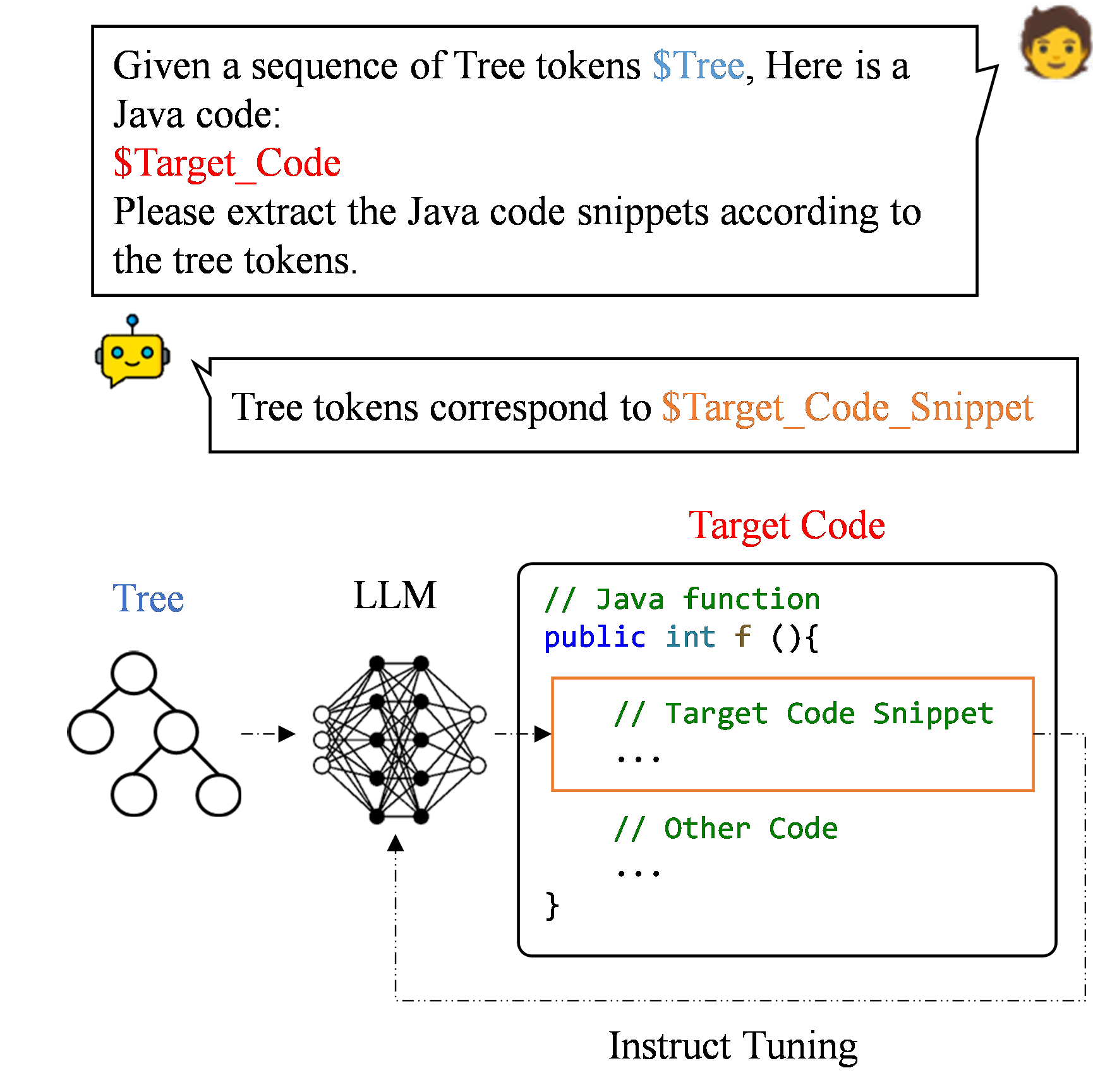}%
\label{fig_sum_2_first_case}}
\hfil
\subfloat[]{\includegraphics[width=3in]{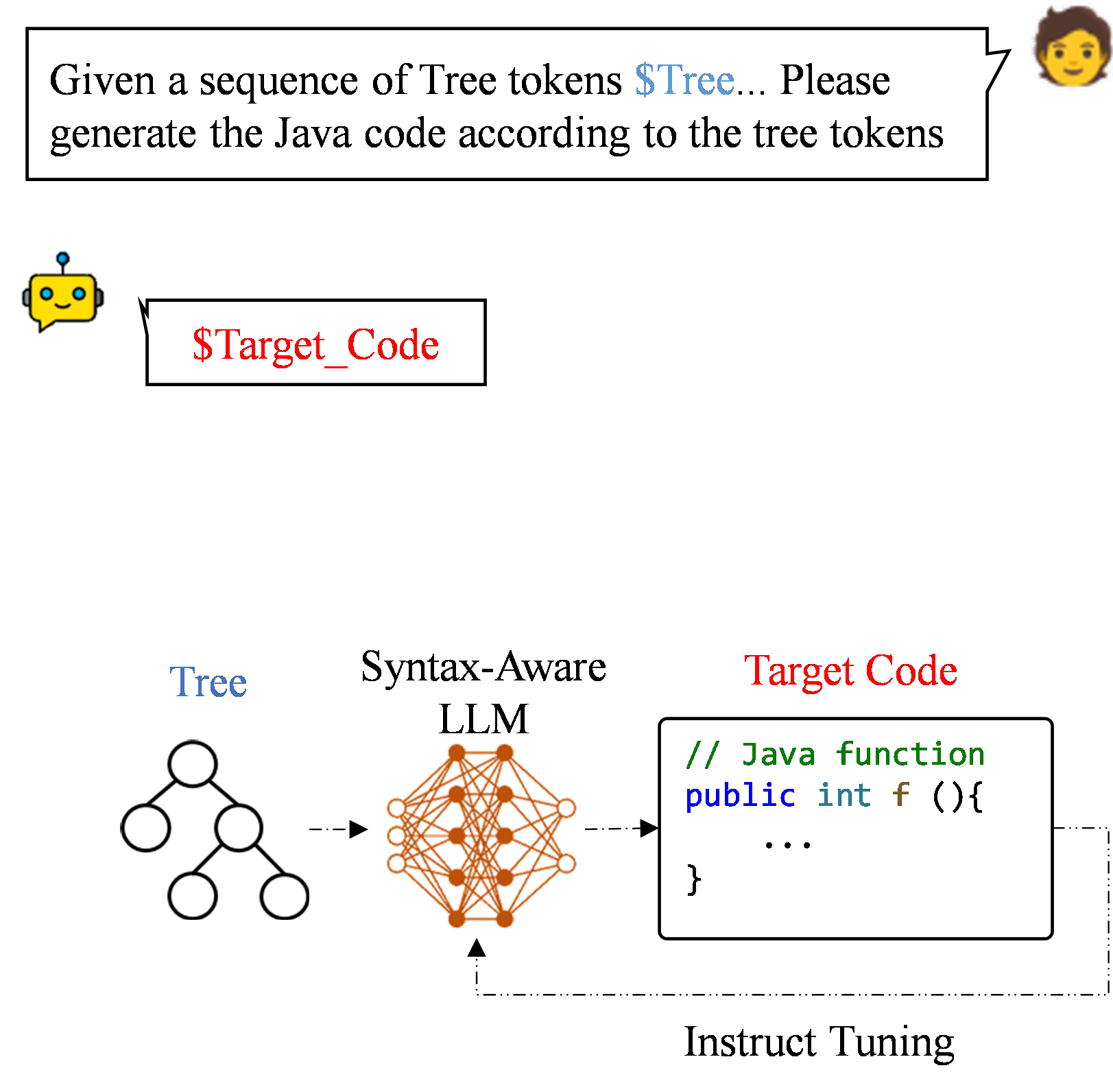}%
\label{fig_sum_2_second_case}}
\caption{The instruction templates of dual-stage tree instruction tuning. (a) Syntax-aware instruction template. (b) Code generation instruction template.}
\label{fig_sum_2}
\end{figure*}

This module expands on foundational work \cite{self-instruction, graphgpt}, employing syntax-aware fine-tuning and code generation fine-tuning to enable LLMs to comprehend domain-specific input structures and downstream tasks. We aim to enable LLMs to learn essential syntactic knowledge from syntactic information representations independently, align tree-structured syntactic features with target code, and generate accurate target code based on these representations.

\textbf{Syntax-Aware Fine-Tuning}. In the first fine-tuning stage, we establish self-supervised fine-tuning using fine-grained parallel datasets to develop a syntax-aware LLM capable of fine-grained syntax-semantic comprehension. This mechanism \cite{instruction_mechanism} captures subtle relations between tree-structural syntactic information representations and code snippets, allowing the LLM to autonomously associate information from tree-nodes to the relevant target code snippets, thereby improving the accuracy of contextual reasoning. At the same time, the integration of fine-grained parallel datasets significantly reduces the LLM's cognitive burden in interpreting complex function-level patterns. Specific instruction templates are shown in Fig. \ref{fig_sum_2}\subref{fig_sum_2_first_case}.

\textbf{Code Generation Fine-Tuning}. In the second fine-tuning stage, we implement supervised fine-tuning to generate complete target code from syntactic information representations. Initialized with the syntax-aware LLM parameters, the model fine-tunes on function-level parallel datasets processed by the syntactic information representation module. This stage results in a code generation LLM capable of generating the corresponding target code directly from the source code's syntactic information representations, while improving executable code generation capability and global alignment accuracy between syntactic structures and target code. Specific instruction templates are shown in Fig. \ref{fig_sum_2}\subref{fig_sum_2_second_case}.

\section{Experimental Setup}
\label{4}

\subsection{Datasets}
\textbf{Statement-Level Parallel Dataset}. We utilize the statement-level parallel dataset from XLCoST to train the matching model in the fine-grained parallel dataset augmentation module. This dataset originates from GeeksForGeeks, a repository containing thousands of data structure and algorithm problems. XLCoST processes these data by segmenting function-level implementations into consecutive code snippets based on code comments, where source-target snippet pairs maintain semantic and functional equivalence.

Table \ref{tab:table2} summarizes the statistical properties of the statement-level parallel code dataset. The original dataset contains solutions across seven programming languages. We select the two most prevalent languages (Python and Java) and remove all comments during preprocessing.

\begin{table}[ht]
\centering
\caption{Statistics of the XLCoST's statement-level parallel dataset.\label{tab:table2}}
\renewcommand{\arraystretch}{1.5}
\begin{tabular}{C{5cm} C{2cm} C{2cm}}
\toprule
\makecell{Translation Pair} & \makecell{Train} & \makecell{Valid} \\
\midrule
Python $\leftrightarrow$ Java & 81,207 & 11,239 \\
\bottomrule
\end{tabular}
\end{table}

\textbf{Function-Level Parallel Datasets}. Within the fine-grained parallel dataset augmentation module, we construct a Python-Java function-level parallel dataset with 20,506 code pairs by filtering four existing resources: (1) AVATAR \cite{avatar}, which is created from computational problem solutions in Java/Python collected from open programming competition platforms; (2) CodeTransOcean \cite{CodeTransOcean}, sourced from a programming chrestomathy website with expert-verified implementations; (3) G-TransEval \cite{G-TransEval}, aggregated from TransCoder-Test and CodeXGLUE \cite{codexglue} benchmarks; (4) XLCoST \cite{xlcost}. These datasets are then used for code generation instruction tuning in the dual-stage tree instruction tuning module.

\begin{table}[ht]
\centering
\caption{Statistics of function-level parallel datasets.\label{tab:table3}}
\renewcommand{\arraystretch}{1.5}
\begin{tabular}{C{3cm}|C{2cm}C{2cm}}
\specialrule{1pt}{0pt}{0pt} 
\makecell{Datasets} & \makecell{Train} & \makecell{Valid} \\
\specialrule{0.8pt}{0pt}{0pt} 
AVATAR         & 6,224 & 1,556 \\
CodeTransOcean & 2,248 & 596   \\
G-TransEval    & 320   & 80    \\
XLCoST         & 8,275 & 2,069 \\
\specialrule{1pt}{0pt}{0pt} 
\end{tabular}
\end{table}

We remove duplicates to improve dataset quality. For each dataset, 20\% of instances are allocated to the valid set, while the remaining are part of the training set. To prevent data leakage, we eliminate samples from the training set with an AST cosine similarity greater than 0.9 with any valid instance. Key statistics of the processed parallel datasets are summarized in Table \ref{tab:table3}.

\textbf{Data for Dual-Stage Tree Instruction Tuning}. We employ the fine-grained parallel dataset augmentation module to create a dataset with 337,602 samples for syntax-aware fine-tuning. For code generation instruction tuning, we leverage function-level parallel datasets to compile a dataset with 20,506 instances.

\begin{table*}[!t]
\caption{Statistics of data for dual-stage tree instruction tuning.\label{tab:table4}}
\centering
\renewcommand{\arraystretch}{2.0} 
\begin{tabular}{C{4cm}|C{2.5cm}C{2.5cm}|C{2.5cm}C{2.5cm}}
\hline
\multirow{2}{*}{Dataset} & 
\multicolumn{2}{c|}{\parbox{5cm}{\centering \vspace{5pt} Syntax-Aware Fine-Tuning \\[5pt] (Fine-Grained Parallel Dataset) \vspace{5pt}}} & 
\multicolumn{2}{c}{\parbox{5cm}{\centering \vspace{5pt} Code Generation Fine-Tuning \\[5pt] (Function-Level Parallel Dataset) \vspace{5pt}}} \\ 

\cline{2-5}
& Train & Valid & Train & Valid \\
\hline
AVATAR & 118,513 & 3,617 & 7,419 & 361 \\
CodeTransOcean & 45,425 & 1,386 & 2,844 & 138 \\
G-TransEval & 6,094 & 185 & 382 & 18 \\
XLCoST & 157,570 & 4,812 & 9,861 & 483\\
\hline
\end{tabular}
\end{table*}

Following \cite{dataset_split}, we partition the syntax-aware fine-tuning dataset by randomly selecting 10,000 instances for validation, with the remaining used for training. For code translation fine-tuning, 1,000 instances are randomly selected for validation, with the rest allocated to training. Detailed dataset statistics are provided in Table \ref{tab:table4}.

\textbf{Test Set}. We evaluate our approach on the multilingual HumanEval-X benchmark, comprising 164 test samples with accompanying unit test cases and reference solutions for semantic and functional correctness verification.

\subsection{Baselines}
In our performance evaluation, we consider the following diverse state-of-the-art approaches for comparison:
\begin{enumerate}
\item{\textbf{Code-specialized LLMs}, including StarCoder2(7B/15B version) and CodeQwen-1.5(7B version);}
\item{\textbf{Specialized code translation models}, with TransCoder and SPT-Code as the representative methods;}
\item{\textbf{General-purpose LLMs}: Given general-purpose LLMs' strong performance in code translation \cite{lost-in-translation}, we include GPT-4 Turbo, DeepSeek-V2(236B version), and DeepSeek-V3 (671B version);}
\item{\textbf{Instruction tuning baselines}: We employ the instruction tuning \cite{IT} as our baseline to demonstrate our approach's advantages over end-to-end fine-tuning in code translation tasks.}
\end{enumerate}

\subsection{Evaluation Metrics}
CodeBLEU \cite{codebleu} is a metric specifically designed for code synthesis tasks. It is calculated as a weighted combination of four components: (1) the original BLEU \cite{bleu} score, (2) weighted n-gram matching, (3) syntactic AST matching, and (4) semantic data-flow matching. The BLEU component reflects identifier naming consistency between translated and source code, while the semantic data-flow score indicates functional equivalence in implementation. We argue that this metric has the potential to evaluate the user-friendliness of translation results.

Our evaluation primarily adopts the metric proposed in \cite{lost-in-translation}, where a code translation is considered successful only if it: (1) compiles without errors, (2) passes all runtime checks, and (3) satisfies the original test cases.

Additionally, we introduce a novel syntactic confusion metric that quantifies the frequency of syntactic confusion occurrences to evaluate how efficiently LLMs comprehend syntactic structures across different programming languages.

\subsection{Implementation Details}
Our implementation leverages PyTorch and the Transformers library. We selected StarCoder2-7B, StarCoder2-15B, and CodeQwen1.5-7B as base models and implemented the TIT method following the configuration of \cite{graphgpt} : per-GPU batch size = 2, learning rate = $2e^{-3}$, warmup ratio = $3e^{-2}$, and joint fine-tuning using LoRA \cite{lora}. For each fine-tuning stage, we set the training epochs to 2.

For evaluating most baselines, we use their publicly available code. Since SPT-Code only released a version for C\#-Java translation tasks, we adapt their approach by training it on a Python-Java function-level parallel dataset to evaluate SPT-Code's performance.

\subsection{Research Questions}
To validate the effectiveness and advantages of TIT, we raised solutions for following research questions (RQs):

\textbf{RQ1: How does TIT perform with various LLMs in code translation task?}

Through architecture-diverse and parameter-scaled fine-tuning with TIT, we evaluate the Python-to-Java code translation performance of fine-tuned LLMs. We compare TIT-integrated LLMs with the state-of-the-art general-purpose LLMs such as GPT-4 Turbo and specialized code translation approaches such as TransCoder to objectively measure TIT's effectiveness in code translation tasks.

\textbf{RQ2: How does TIT’s performance compare to standard instruction tuning (IT) in code translation tasks?}
We implemented TIT and IT on LLMs with various parameter scales and architectures, evaluating their comparative performance in code translation task.

\textbf{RQ3: How effectively does TIT mitigate syntactic confusion in code translation?}
We quantified and comparatively analyzed the incidence of translation failures caused by syntactic confusion during inference across TIT-integrated, IT-integrated, and prototype LLMs to validate TIT's efficacy in mitigating syntactic confusion for code translation tasks.

\textbf{RQ4: How do individual components of TIT contribute to improving code translation performance?}
We performed ablation studies to validate the contributions of the fine-grained parallel dataset augmentation module and the dual-stage tree-structured instruction tuning module to code translation. 

\textit{1) Syntactic Information Representation Module:} To evaluate the contribution of the syntactic information representation module, we directly applied dual-stage tree instruction tuning to three base models using parallel code pairs from the fine-grained and function-level parallel datasets, while adjusting the instructions. For the first-stage fine-tuning, the instruction was modified to: "Given a Python code snippet \$Source\_Code and a Java code snippet \$Target\_Code, extract the Java code snippet based on the Python code snippet." For the second-stage fine-tuning, the instruction was modified to: "Please translate the given Python code \$Source\_Code to Java." We evaluate the fine-tuned models on the HumanEval-X benchmark.

\textit{2) Fine-Grained Parallel Dataset Augmentation Module:} We evaluate the contribution of the fine-grained parallel dataset augmentation module through dual perspectives: data scaling and granularity refinement.

\textbf{Data Scaling}. To evaluate the contribution of data scale in fine-grained parallel datasets, this experiment performs a comparative analysis within the TIT using XLCoST's statement-level dataset as the baseline. Specifically, we substitute the fine-grained parallel dataset with XLCoST's statement-level data for syntax-aware fine-tuning, then incorporate all function-level datasets for code generation fine-tuning. We quantify translation performance using HumanEval-X metrics and benchmark these results against the fine-grained dataset configuration.

\textbf{Granularity Refinement}. We conducted phased incremental testing to validate the performance gains attributable to the granularity refinement of the fine-grained dataset. Specifically, we formulated mixtures with different ratios of fine-grained parallel data (0\%/30\%/50\%/70\%/100\%) and complementary original function-level data (100\%/70\%/50\%/30\%/0\%). These mixtures were used for syntax-aware fine-tuning of the base model. Subsequently, we performed unified fine-tuning for code generation on the syntax-aware LLMs using the complete function-level dataset. Finally, we evaluated the improvements in translation performance based on the HumanEval-X benchmark.

\textit{3) Dual-Stage Tree Instruction Tuning Module:}To analyze the contribution of the dual-stage tree instruction tuning module, we directly applied code generation fine-tuning to the base model using the function-level parallel dataset processed by the syntactic information representation module. The performance of the resulting code generation LLMs was then evaluated on the test set.

\section{Results \& Analysis}

\subsection{RQ1: How does TIT perform with various LLMs in code translation task?}

\begin{figure}[!t]
\centering
\includegraphics[width=5in]{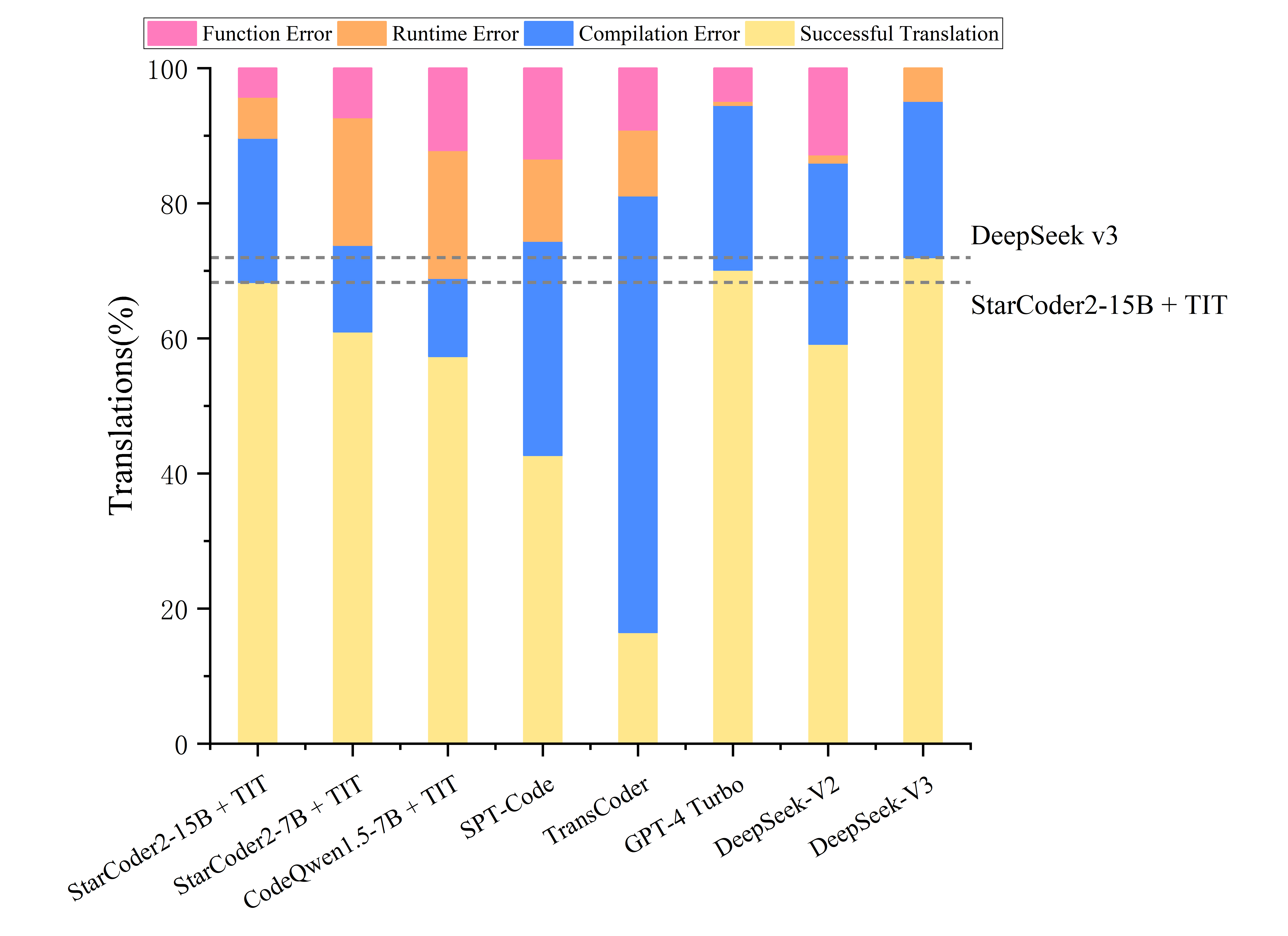}
\caption{ Performance of the model with TIT applied compared to the baseline in P2J translation.}
\label{fig_9}
\end{figure}

\begin{figure}[!t]
\centering
\includegraphics[width=5in]{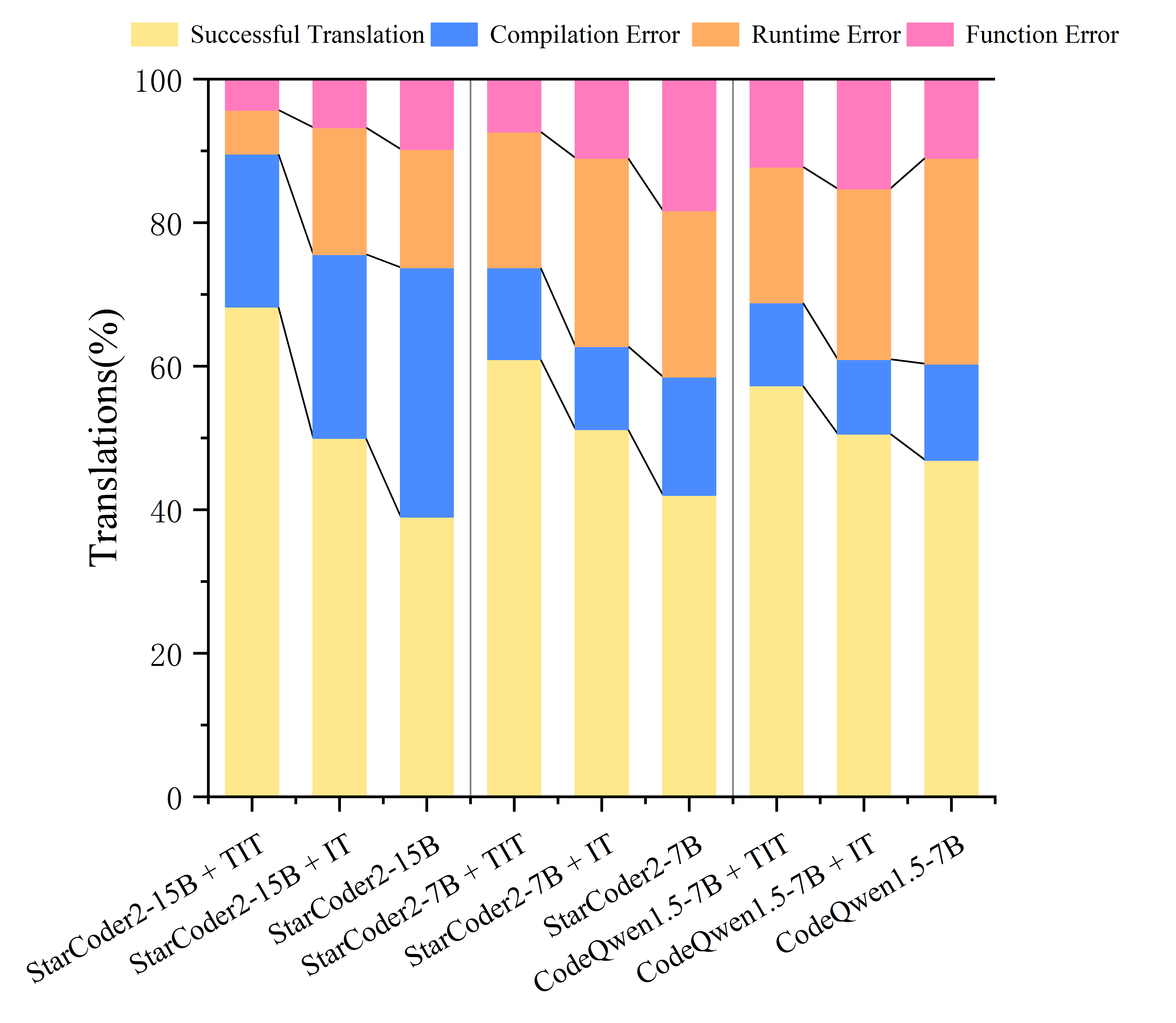}
\caption{ Performance of the model with TIT applied compared to the base model in P2J translation.}
\label{fig_10}
\end{figure}

\begin{table}[ht]
\centering
\caption{Results of successful translation rate and CodeBLEU metrics in P2J translation.\label{tab:table4}}
\renewcommand{\arraystretch}{1.5}
\begin{tabular}{C{5cm}|C{2.5cm}C{2cm}}
\specialrule{1pt}{0pt}{0pt} 
Approach & Successful Rate (\%) & CodeBLEU (\%) \\
\specialrule{0.8pt}{0pt}{0pt} 
StarCoder2-15B & 39.02 & 56.43 \\
StarCoder2-7B & 42.07 & 35.44 \\
CodeQwen1.5-7B & 46.95 & 61.17 \\
SPT-Code & 42.68 & 60.53 \\
TransCoder & 16.46 & 58.90 \\
GPT 4-Turbo & 70.12 & 59.69 \\
DeepSeek-V2 & 59.15 & 62.95 \\
DeepSeek-V3 & 71.95 & 65.22 \\
\hline
\textbf{StarCoder2-15B + TIT} & \textbf{68.29} & \textbf{62.06} \\
\textbf{StarCoder2-7B + TIT} & \textbf{60.98} & \textbf{59.81}   \\
\textbf{CodeQwen1.5-7B + TIT} & \textbf{57.32} & \textbf{67.12}    \\
\specialrule{1pt}{0pt}{0pt} 
\end{tabular}
\end{table}

Fig. \ref{fig_9}, Fig. \ref{fig_10} and Table \ref{tab:table4} present comparative translation success rates of our approach versus state-of-the-art models in Python-to-Java translation task. 

Our method achieved translation success rates of 68.9\%, 60.98\%, and 57.32\% on StarCoder2-15B, StarCoder2-7B, and CodeQwen1.5-7B, respectively—showing performance improvements of 1.75×, 1.45×, and 1.22× over their respective base models. StarCoder2-15B derived the greatest benefit from TIT, attributable to its parameter scale advantage, where larger LLMs more effectively leverage syntactic information representations in TIT. 

Compared to specialized code translation models, our approach achieves state-of-the-art results. TransCoder's significant performance difference confirms the limitations of purely statistical machine translation methods and monolingual training in bridging cross-linguistic syntactic divides. TIT elevates the translation success rate of base models to 3.48×–4.15× that of TransCoder. While SPT-Code outperforms TransCoder by incorporating syntactic structures, its lexical-level alignment fails to overcome model parameter limitations. Our method consistently improves success rates by 1.34×–1.60× over SPT-Code.

Compared to general-purpose LLMs, the StarCoder2-15B model integrated with TIT achieves performance close to that of GPT-4 Turbo (70.12\%). Despite its significantly smaller parameter count, its translation capability reaches 97.4\% of GPT-4 Turbo's performance. This powerfully demonstrates the effectiveness of TIT in mitigating the performance gap associated with model scaling. DeepSeek-V3 (71.95\%) achieved the highest translation success rate of 71.95\%, attributable to its architectural advantages. Notably, our best result maintains a 1.15× advantage over DeepSeek-V2 (59.15\%) while utilizing a base model with a significantly smaller parameter count.

As shown in Table \ref{tab:table4}, our method achieves CodeBLEU scores comparable to those of larger LLMs. The TIT-integrated CodeQwen1.5-7B delivers the best performance at 67.12\%. Compared to their base models, TIT improves the CodeBLEU score by 1.10× to 1.69× on base models. This demonstrates that our method effectively preserves functional implementations and identifier naming conventions from the source code in the translated output. Notably, while SPT-Code and TransCoder exhibit strong performance on CodeBLEU (50.90\% and 60.53\%), their overfitting to these metrics fails to translate into significant gains in translation success rate. This reveals a significant divergence between optimizing for static metrics and achieving real improvements in executability.

\subsection{RQ2: How does TIT’s performance compare to standard instruction tuning (IT) in code translation tasks?}

Figure \ref{fig_10} and Table \ref{tab:table4} present the translation success rates when deploying TIT and IT within base models. All model configurations exhibit the performance hierarchy: Prototype \textless IT \textless TIT, demonstrating TIT’s substantive contribution to translation success rates. Under identical data provisioning, IT enhances translation success rates by factors of 1.28×, 1.21×, and 1.08× for StarCoder2-15B, StarCoder2-7B, and CodeQwen1.5-7B, respectively. Our method further outperforms IT by factors ranging from 1.13× to 1.37×. 

\begin{figure}[!t]
\centering
\includegraphics[width=5in]{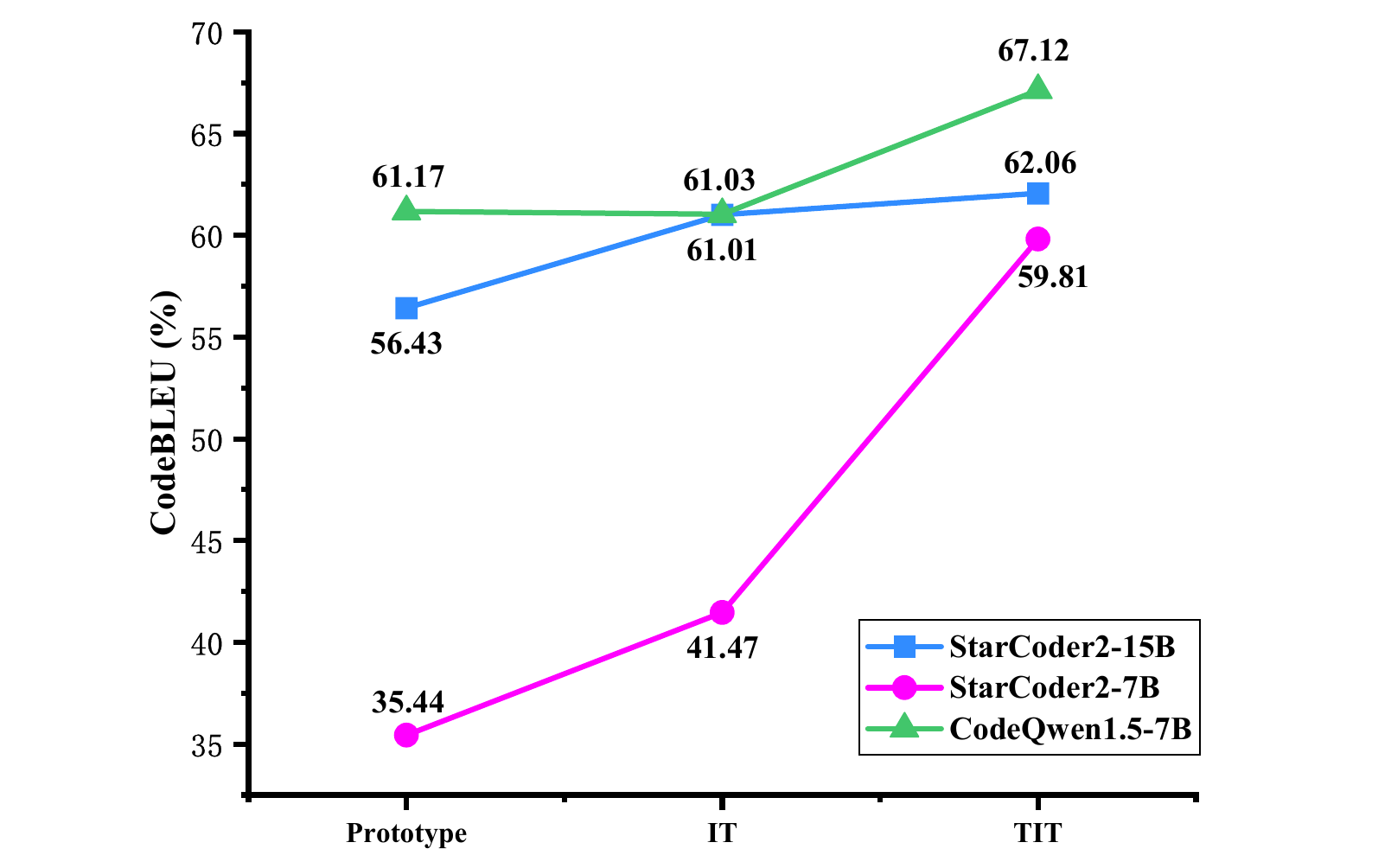}
\caption{ Performance of TIT and IT in the CodeBLEU metric.}
\label{fig_11}
\end{figure}

Table \ref{tab:table4} and Figure \ref{fig_11} show the CodeBLEU metric results from deploying TIT and IT within base models. Overall, TIT improves all CodeBLEU metrics. This improvement stems from CodeBLEU's sensitivity to syntactic matching, where syntax-aware training in TIT enhances metric scores by better maintaining functional implementation and identifier naming fidelity from source code. In contrast, IT provides only marginal CodeBLEU gains and even exhibits slight degradation in CodeQwen1.5-7B. This is attributable to insufficiently precise syntactic information representations.

These results collectively show that using syntactic information representations as an intermediate language offers greater potential for code translation task than end-to-end direct IT approaches.

\subsection{RQ3: How effectively does TIT mitigate syntactic confusion in code translation?}

\begin{figure}[!t]
\centering
\includegraphics[width=5in]{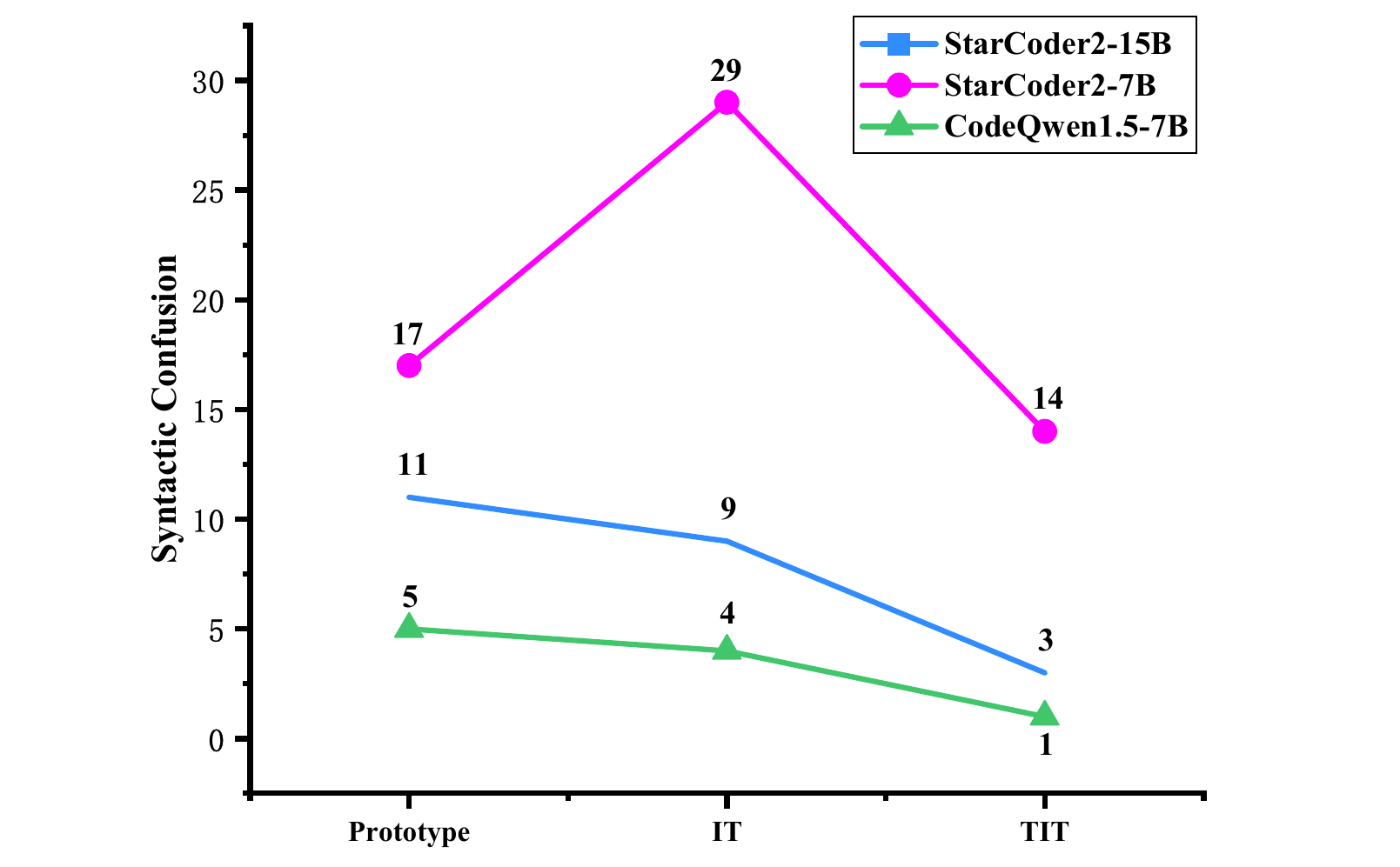}
\caption{Performance of TIT for LLM syntactic confusion improvement.}
\label{fig_12}
\end{figure}

The results are demonstrated in Fig. \ref{fig_12}. Our approach significantly mitigates syntactic confusion across all base models, reducing rates by 72.73\% for StarCoder2-15B, 17.65\% for StarCoder2-7B, and 80.00\% for CodeQwen-7B. This improvement stems from excluding language-specific syntax in the inputs, thereby minimizing redundant information that could mislead LLMs. The variation in effectiveness across models reflects differential capabilities in processing syntactic representations based on parameter scale and architecture.

Conversely, the IT introduces language-specific tokens without refining source-language information. This ambiguity causes LLMs to conflate translation with code completion tasks, substantially compromising output quality. Most notably, IT increased compilation errors from syntactic confusion in StarCoder2-7B (29 errors vs. 17 in the base model). This suggests that direct instruction tuning without input refinement may adversely impact translation performance, particularly in smaller-scale models.

\subsection{RQ 4: How do individual components of TIT contribute to enhancing code translation performance?}

\begin{figure}[!t]
\centering
\subfloat[]{\includegraphics[width=3in]{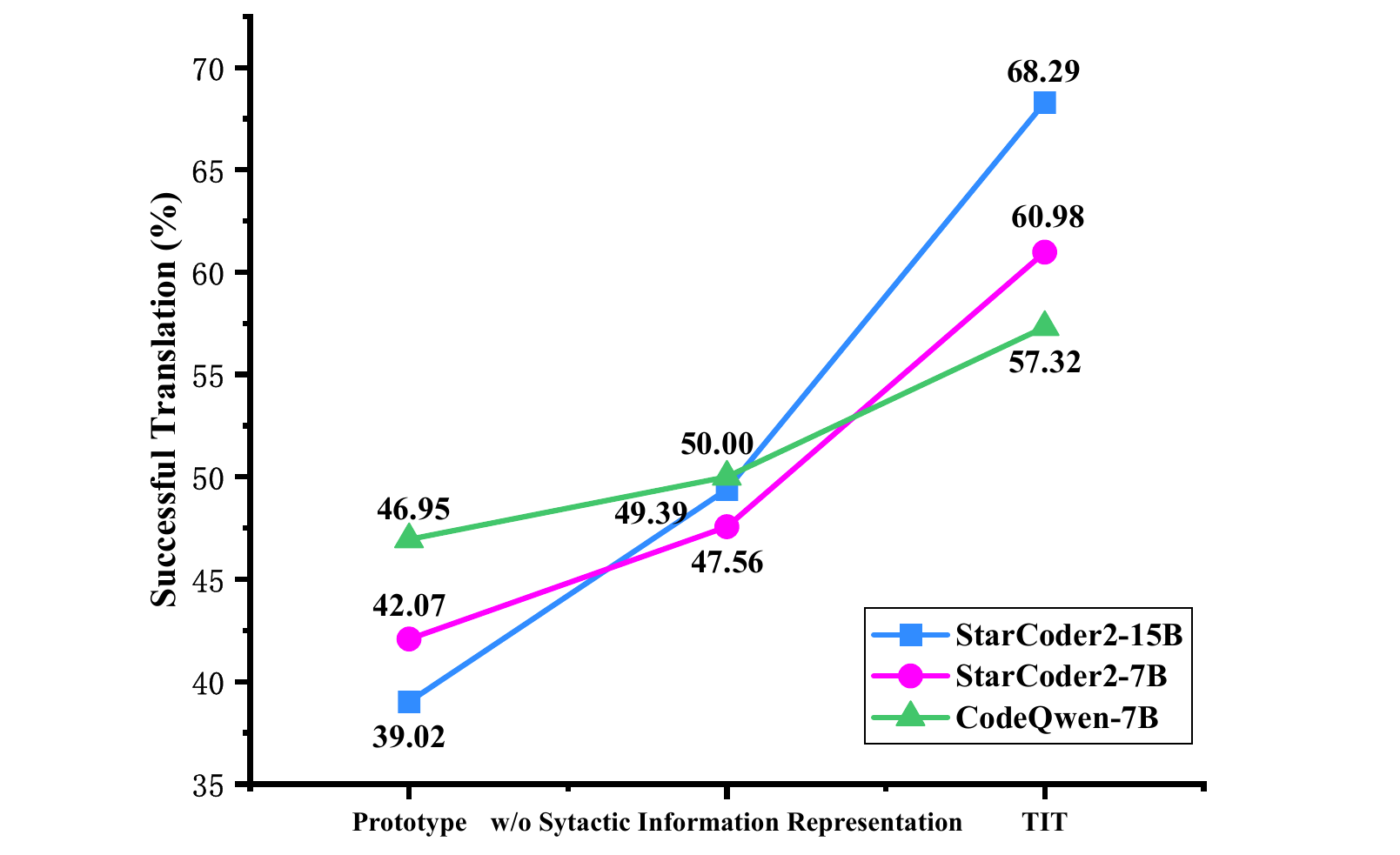}%
\label{fig_sum_6_first_case}}
\hfil
\subfloat[]{\includegraphics[width=3in]{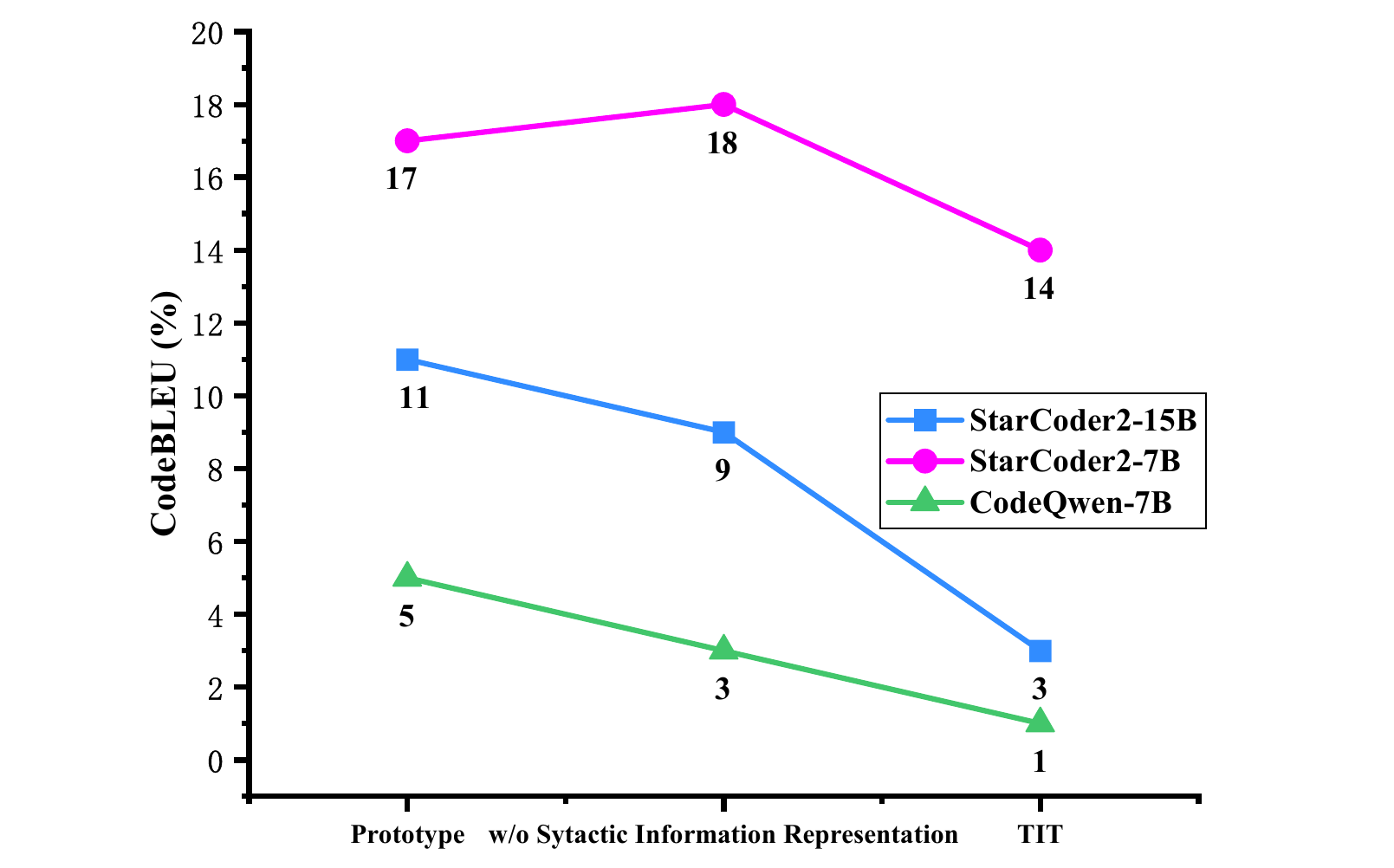}%
\label{fig_sum_6_second_case}}
\hfil
\subfloat[]{\includegraphics[width=3in]{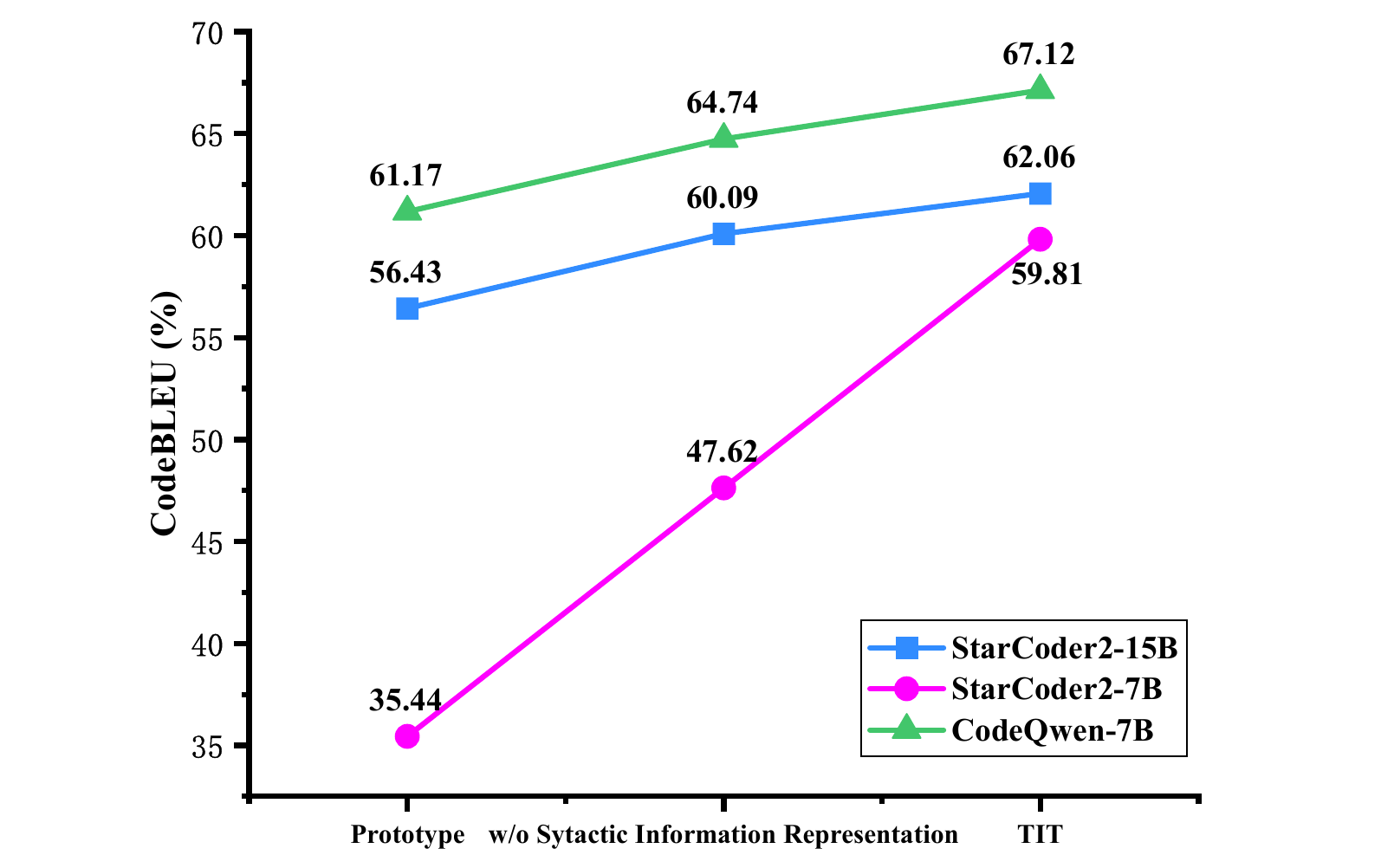}%
\label{fig_sum_6_third_case}}
\caption{The performance of TIT whether using the Syntactic Information Representation Module. (a) Successful translation rate. (b) Syntactic confusion. (c) CodeBLEU.}
\label{fig_sum_6}
\end{figure}

\textit{1) Syntactic Information Representation Module:} Fig. \ref{fig_sum_6} illustrates the performance of different model configurations across various metrics after removing the syntactic information representation module.

Fig. \ref{fig_sum_6}\subref{fig_sum_6_first_case} reports the code translation success rates under different model configurations. Compared to the prototype model, removing the syntactic information representation module (w/o syntactic information representation module) improves the success rates of StarCoder2-15B, StarCoder2-7B, and CodeQwen-7B by 26.56\%, 13.04\%, and 6.49\%, respectively. This indicates that fine-grained parallel data and optimized instruction tuning are the primary drivers of performance improvement.

Further incorporating the syntactic information representation module yields substantial gains across all base models. Specifically, StarCoder2-15B achieves an additional 38.27\% improvement over w/o syntactic information representation module, corresponding to a 75.00\% overall gain relative to the prototype model. StarCoder2-7B improves by 28.21\%, leading to an overall increase of 44.93\%, while CodeQwen-7B shows a further 14.63\% improvement, reaching a total gain of 22.08\% over the prototype model.

Fig. \ref{fig_sum_6}\subref{fig_sum_6_first_case} reports the impact of different module combinations on code translation success rates. Compared to the Prototype, removing the syntactic information representation module (w/o syntactic information representation module) improved the success rate of StarCoder2-15B, StarCoder2-7B, and CodeQwen-7B by 26.56\%, 13.04\%, and 6.49\%, respectively. This indicates that fine-grained parallel data augmentation and dual-stage tree instruction tuning are the key factors driving performance improvements. With the further introduction of the syntactic information representation module, performance gains are observed across all backbone models. Specifically, StarCoder2-15B achieved an additional 38.27\% improvement over w/o syntactic information representation module, leading to an overall 75.00\% increase compared to the Prototype. StarCoder2-7B improved by 28.21\% (overall 44.93\%), while CodeQwen-7B improved by 14.63\% (overall 22.08\%). These results clearly demonstrate that the syntactic information representation module can further enhance translation accuracy, especially in large-scale LLMs.

Fig. \ref{fig_sum_6}\subref{fig_sum_6_second_case} presents the results of syntactic confusion errors across different model configurations. Compared with the Prototype, removing the syntactic information representation slightly alleviates this issue, reducing confusion errors from 11 to 9 in StarCoder2-15B and from 5 to 3 in CodeQwen-7B, though StarCoder2-7B shows a minor increase (from 17 to 18). These results suggest that fine-grained data and instruction tuning help mitigate confusion but are insufficient for all models. With the further introduction of the syntactic information representation module, syntactic confusion is significantly reduced across all base models. Specifically, the confusion decreases by 66.67\% for StarCoder2-15B, 22.22\% for StarCoder2-7B, and 66.67\% for CodeQwen-7B. These results indicate that the syntactic information representation module effectively helps LLMs distinguish between code continuation and code translation tasks, thereby mitigating syntactic confusion in the generated output.

For the CodeBLEU metric, results are shown in Fig. \ref{fig_sum_6}\subref{fig_sum_6_third_case}. Compared to the Prototype, removing the syntactic information representation module improved StarCoder2-15B, StarCoder2-7B, and CodeQwen-7B by 6.48\%, 34.39\%, and 5.83\%, respectively. With the introduction of the syntactic information representation module, StarCoder2-15B achieved an additional 3.28\% improvement, reaching an overall 9.97\% increase. StarCoder2-7B improved by 25.66\% while CodeQwen-7B improved by 3.67\%. These findings indicate that syntactic information representation further strengthens syntactic and semantic consistency, thereby improving the usability and functional reliability of translated code.

This ablation study demonstrates that the fine-grained parallel dataset and the dual-stage tree instruction tuning module are the primary drivers of translation performance. In addition, the syntactic information representation module further alleviates syntactic confusion, enhances semantic alignment, and provides significant gains for LLMs.

\textit{2) Fine-Grained Parallel Dataset Augmentation Module:} To assess how detailed parallel datasets impact code translation, we compare them with XLCoST’s statement-level dataset and function-level datasets. 
\begin{figure}[!t]
\centering
\subfloat[]{\includegraphics[width=3in]{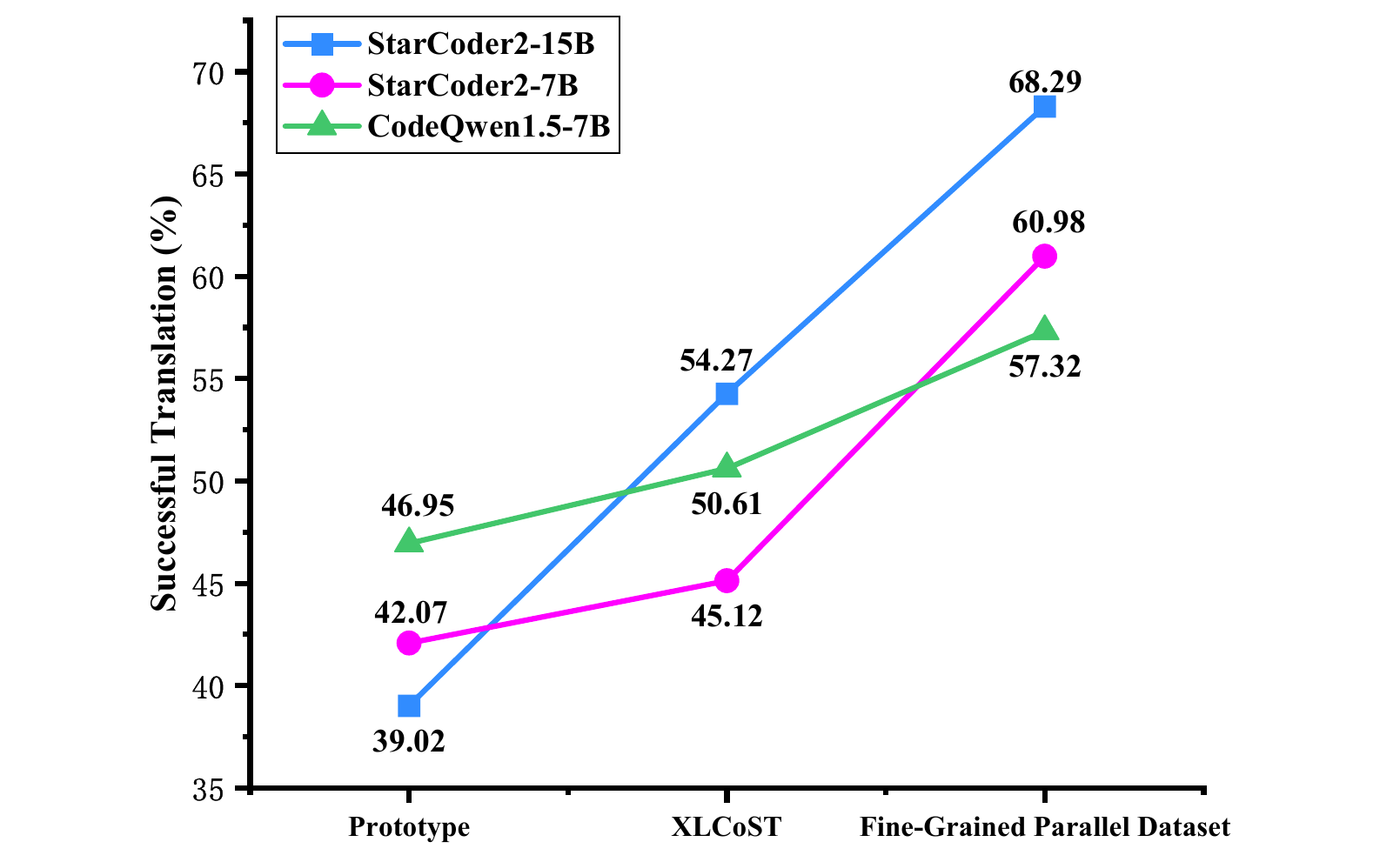}%
\label{fig_sum_3_first_case}}
\hfil
\subfloat[]{\includegraphics[width=3in]{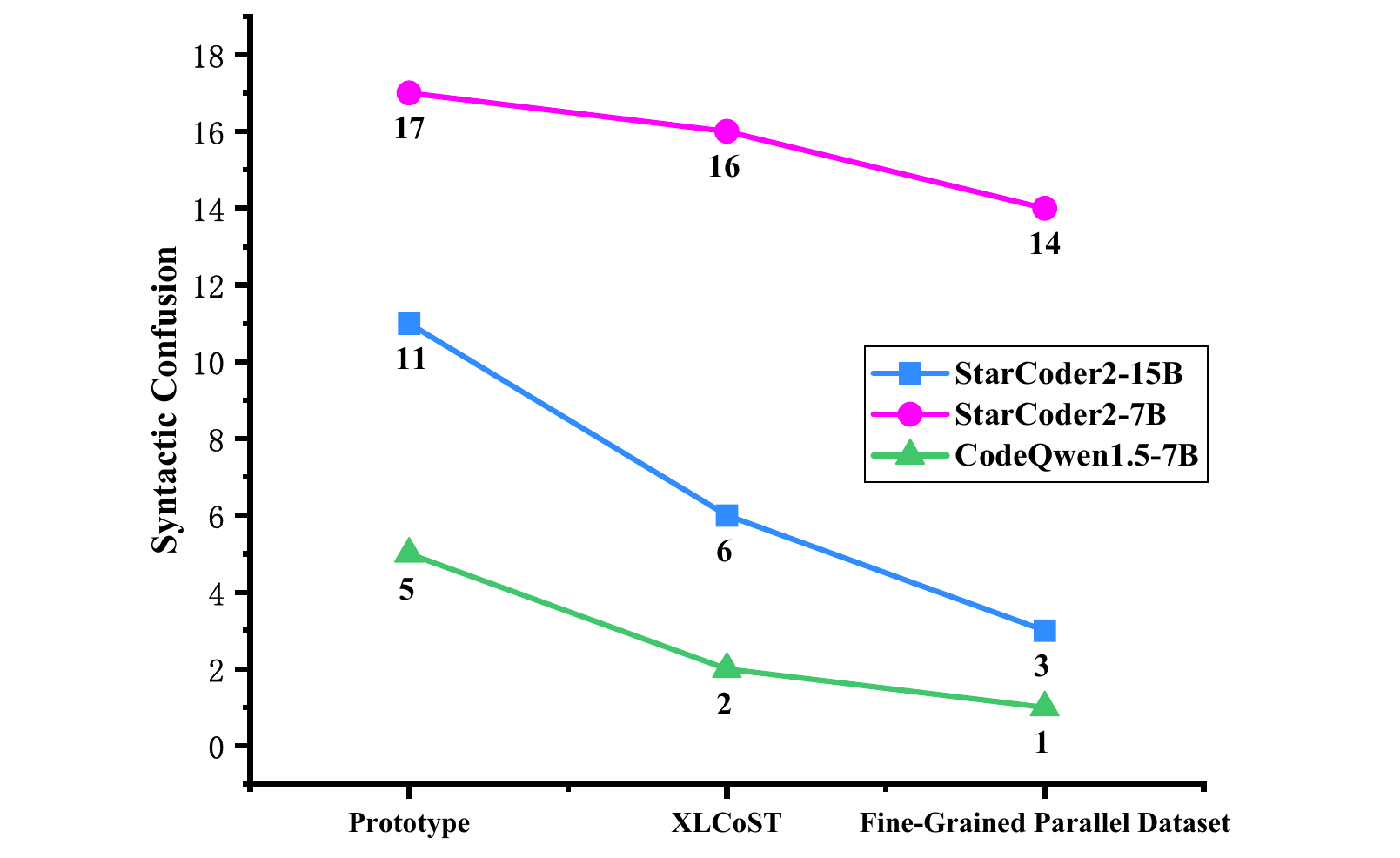}%
\label{fig_sum_3_second_case}}
\hfil
\subfloat[]{\includegraphics[width=3in]{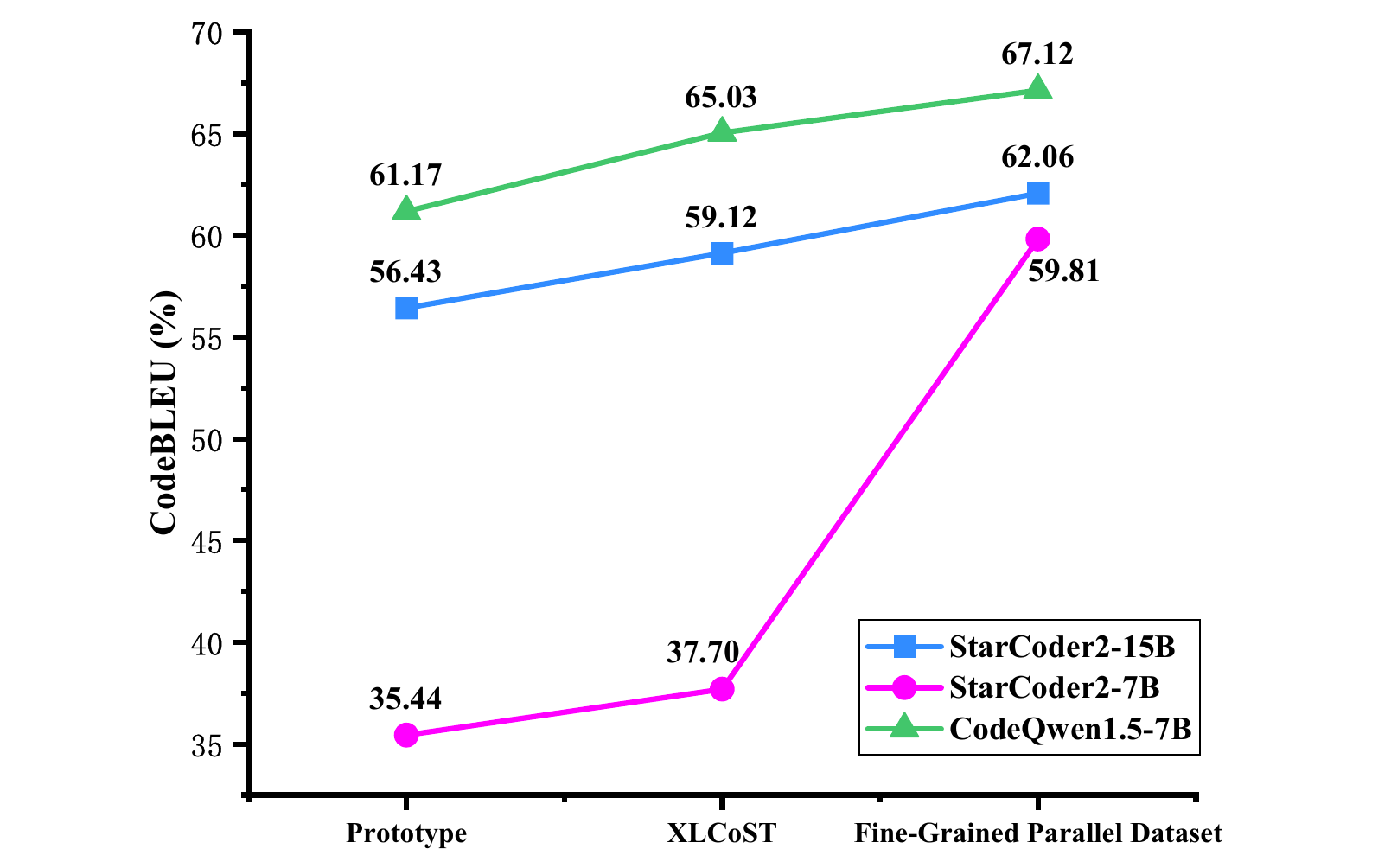}%
\label{fig_sum_3_third_case}}
\caption{Data scaling benefits from using fine-grained parallel datasets. (a) Successful translation rate. (b) Syntactic confusion. (c) CodeBLEU.}
\label{fig_sum_3}
\end{figure}

\textbf{Data Scaling}. Fig. \ref{fig_sum_3} demonstrates the data scaling benefits from using fine-grained parallel datasets. Experimental results indicate that all LLMs perform better across metrics with the proposed fine-grained parallel dataset (about 3.65× larger than XLCoST) compared to XLCoST. Specifically, in terms of code translation success rate, the fine-grained parallel dataset improves results over XLCoST by 1.26×, 1.45×, and 1.22× for StarCoder2-15B, StarCoder2-7B, and CodeQwen1.5-7B, respectively, which can be attributed to the fact that the fine-grained dataset contains approximately 3.80× more samples than XLCoST. Additionally, on the syntactic confusion metric, it reduces errors by 1.60×, 3.00×, and 1.33× across these models. The fine-grained dataset also consistently outperforms XLCoST on the CodeBLEU metric for all baseline models. These results substantiate the contribution of increased data scale in the fine-grained parallel dataset to code translation performance. Additionally, the performance gains observed with the XLCoST dataset on baseline models reinforce the value of fine-grained data.

\begin{figure}[!t]
\centering
\subfloat[]{\includegraphics[width=3in]{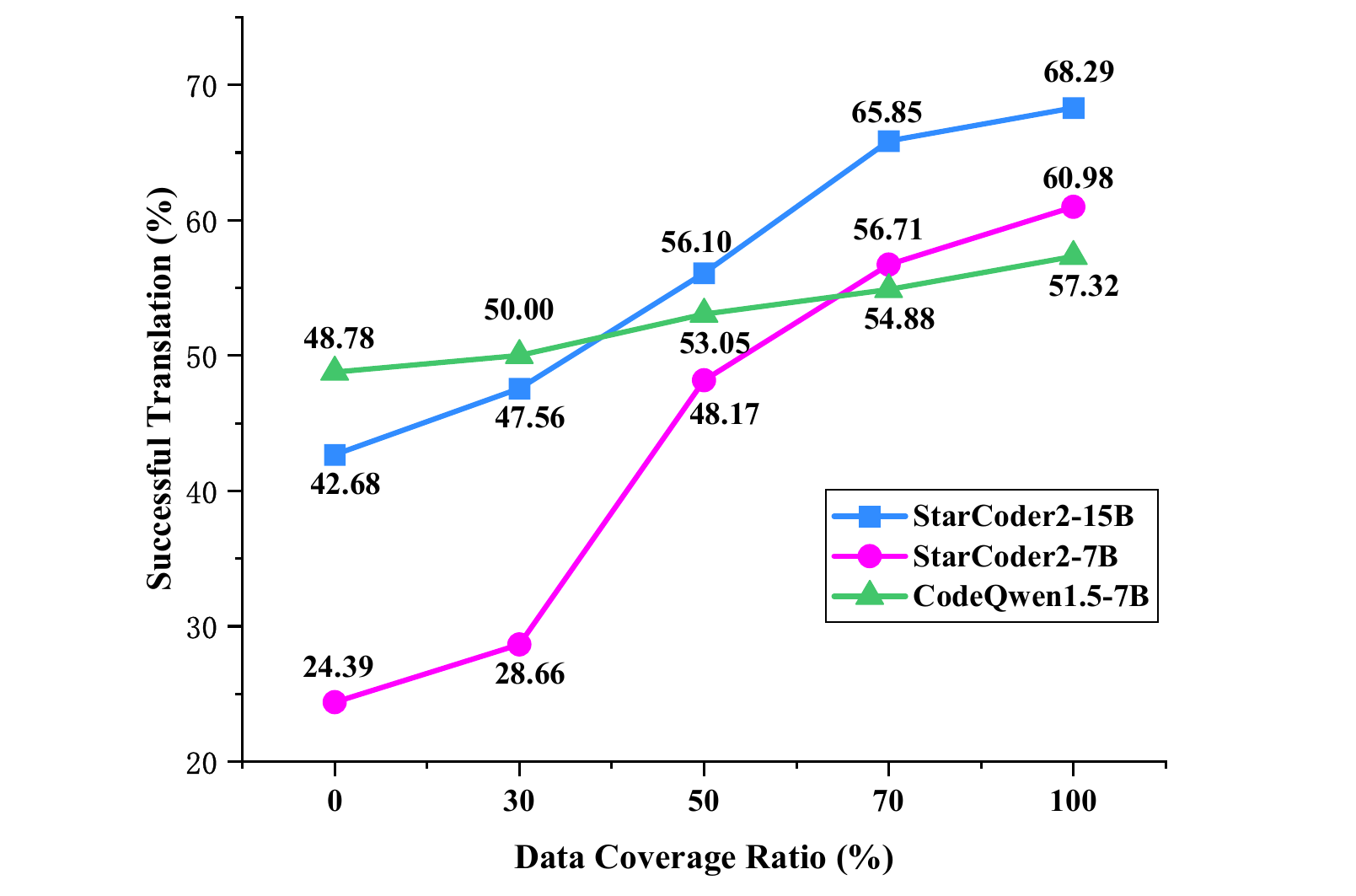}%
\label{fig_sum_4_first_case}}
\hfil
\subfloat[]{\includegraphics[width=3in]{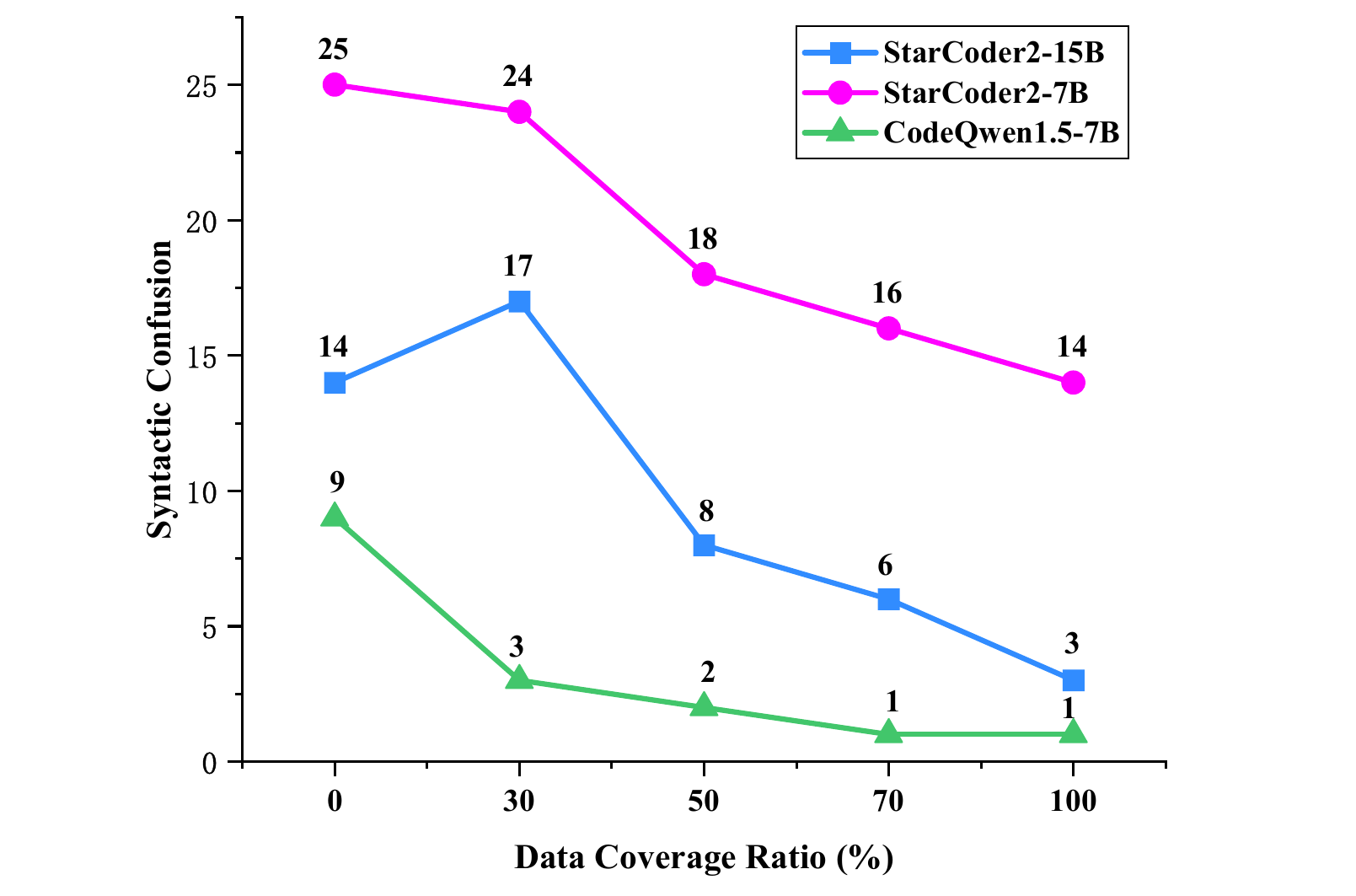}%
\label{fig_sum_4_second_case}}
\hfil
\subfloat[]{\includegraphics[width=3in]{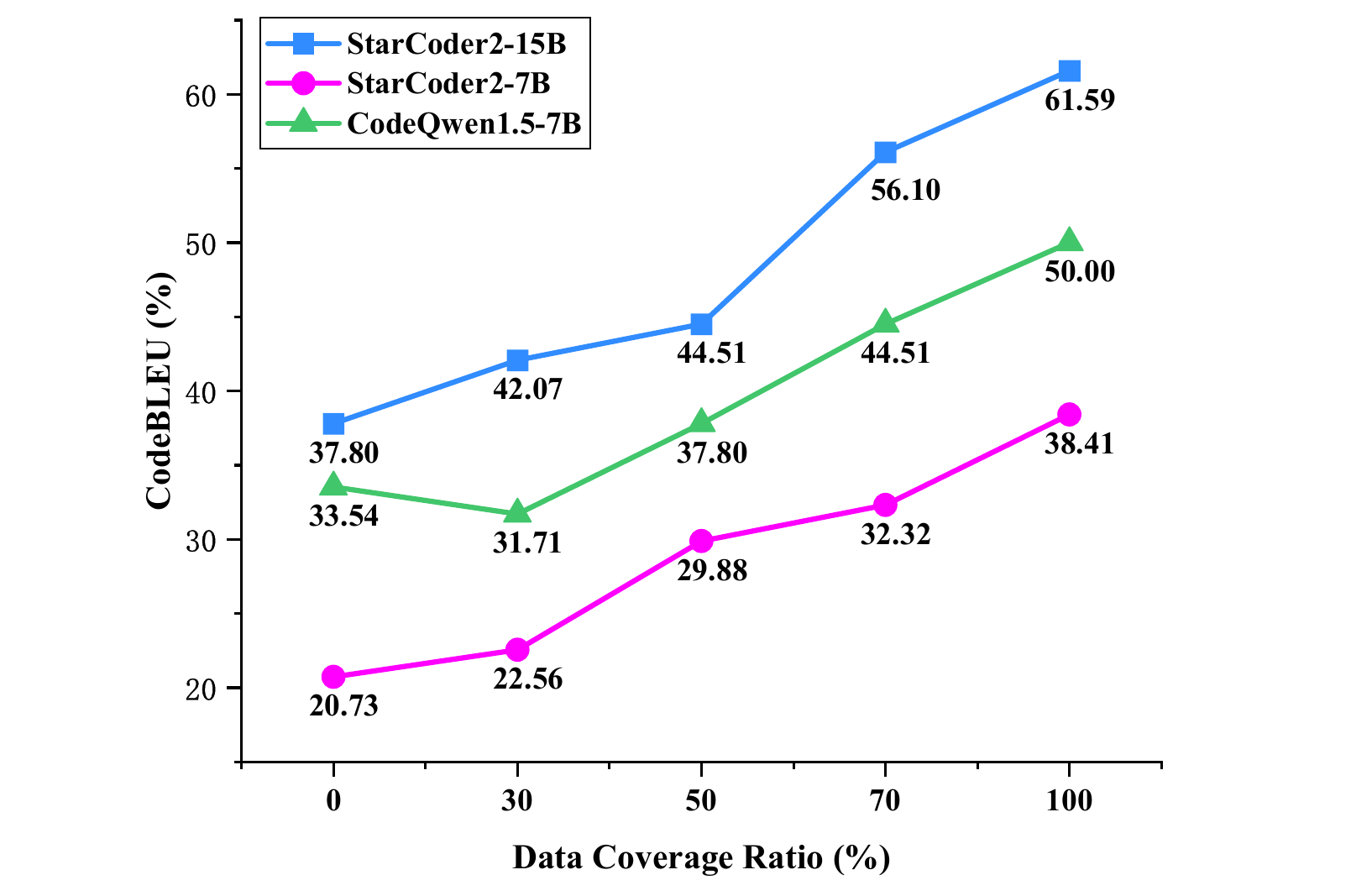}%
\label{fig_sum_4_third_case}}
\caption{Performance improvements from granularity refinement in fine-grained parallel datasets. (a) Successful translation rate. (b) Syntactic confusion. (c) CodeBLEU.}
\label{fig_sum_4}
\end{figure}

\textbf{Granularity Refinement}. Fig. \ref{fig_sum_4} present the performance improvements achieved through granularity refinement in fine-grained parallel datasets.

Fig. \ref{fig_sum_4}\subref{fig_sum_4_first_case} shows the success rates of code translation across varying ratios of fine-grained parallel dataset (0\%–100\%). The integration of these datasets yields significant performance improvements for the StarCoder2 series models, with a turning point at 30\% data, maximum benefits at 50\%, and diminishing returns beyond 70\%. Meanwhile, CodeQwen1.5-7B shows steady improvements throughout the scaling process.

Fig. \ref{fig_sum_4}\subref{fig_sum_4_second_case}  demonstrates the contribution of fine-grained parallel datasets at varying ratios toward mitigating syntactic confusion. During the 0\%-30\% data ratio coverage of fine-grained parallel dataset, model performance varied significantly due to parametric and architectural differences: CodeQwen1.5-7B achieved substantial reduction (-66.70\%) while StarCoder2-7B showed minimal change (-4.00\%), and StarCoder2-15B exhibited performance fluctuation (+21.4\%) when processing high-quality small-batch data. The 30\%-50\% data ratio coverage featured efficient correction across all models with a mean reduction of 37.2\%, where StarCoder2-15B demonstrated the strongest syntax error mitigation (-52.90\%). During the 50\%-100\% data ratio coverage, marginal returns diminished significantly across all models, yielding a mean reduction of only 15.3\%. 

Fig. \ref{fig_sum_4}\subref{fig_sum_4_third_case}  shows CodeBLEU results across varying data ratios of fine-grained parallel data. Our experiments indicate that augmenting with fine-grained data significantly improves semantic consistency in code translation. At 50\% data ratio coverage, CodeQwen1.5-7B exhibits a superlinear growth inflection with 31.00\% CodeBLEU improvement, revealing its exceptional capacity to capture AST features and data-flow dependencies through statement-level alignment. When coverage reaches 100\%, it achieves a CodeBLEU score of 67.12\%, which is a 144.80\% increase over 0\% coverage, confirming the method's effectiveness in activating cross-representational alignment potential within this architecture.

These findings indicate that while parameter scale and architecture modulate LLM performance, all models gain benefits in logical equivalence and syntactic matching benefits from granular refinement in fine-grained parallel datasets.

\begin{figure}[!t]
\centering
\subfloat[]{\includegraphics[width=3in]{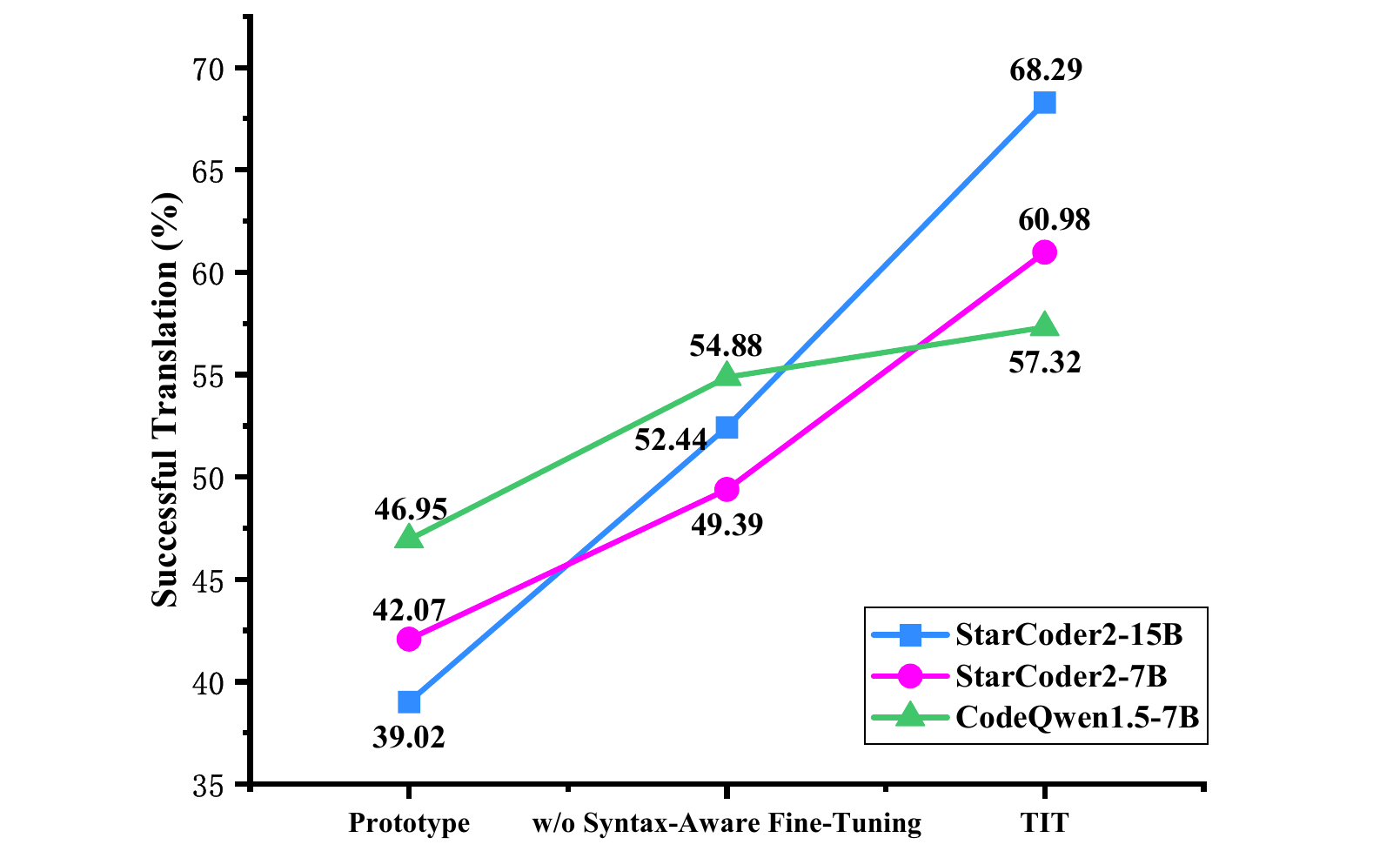}%
\label{fig_sum_5_first_case}}
\hfil
\subfloat[]{\includegraphics[width=3in]{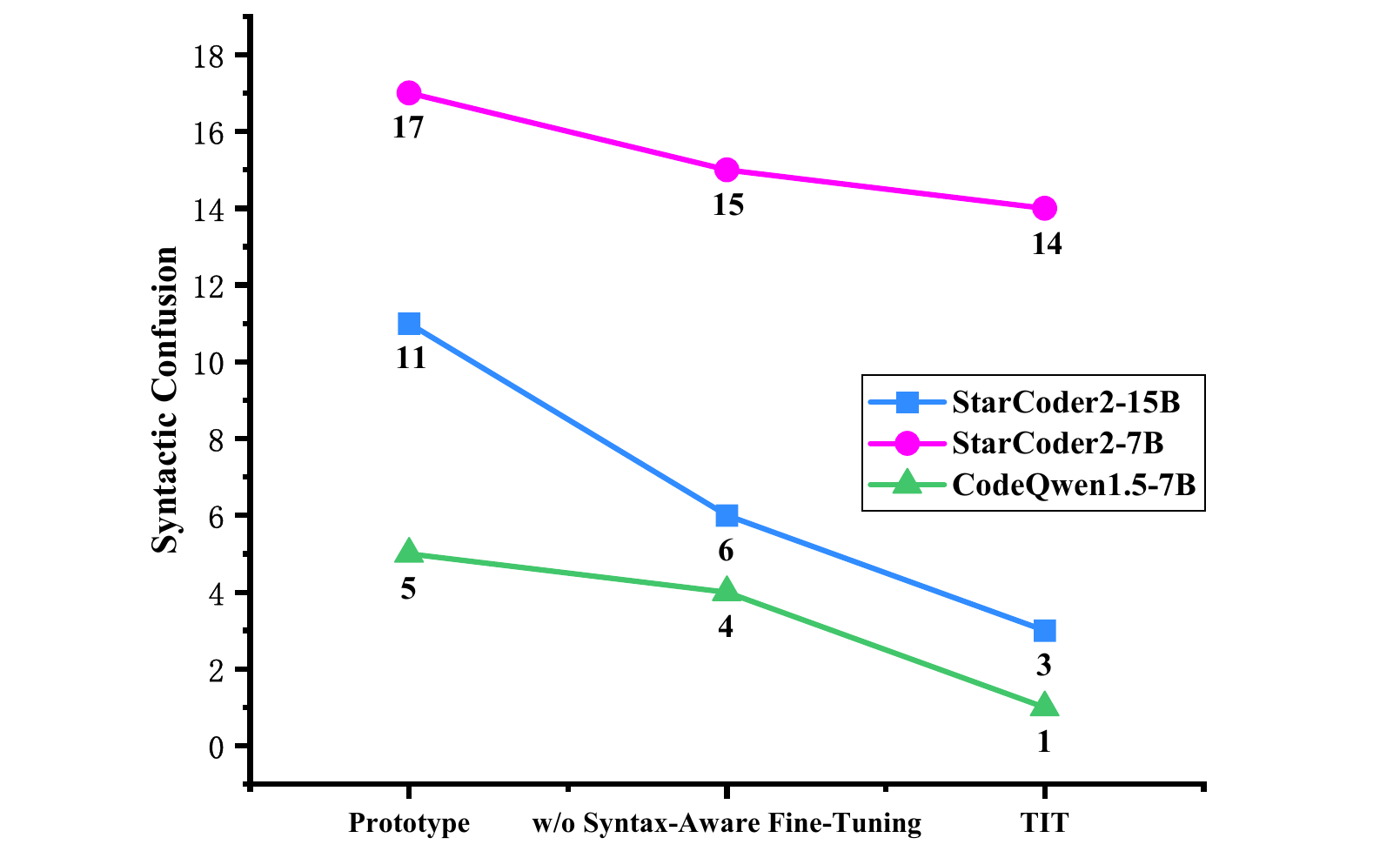}%
\label{fig_sum_5_second_case}}
\hfil
\subfloat[]{\includegraphics[width=3in]{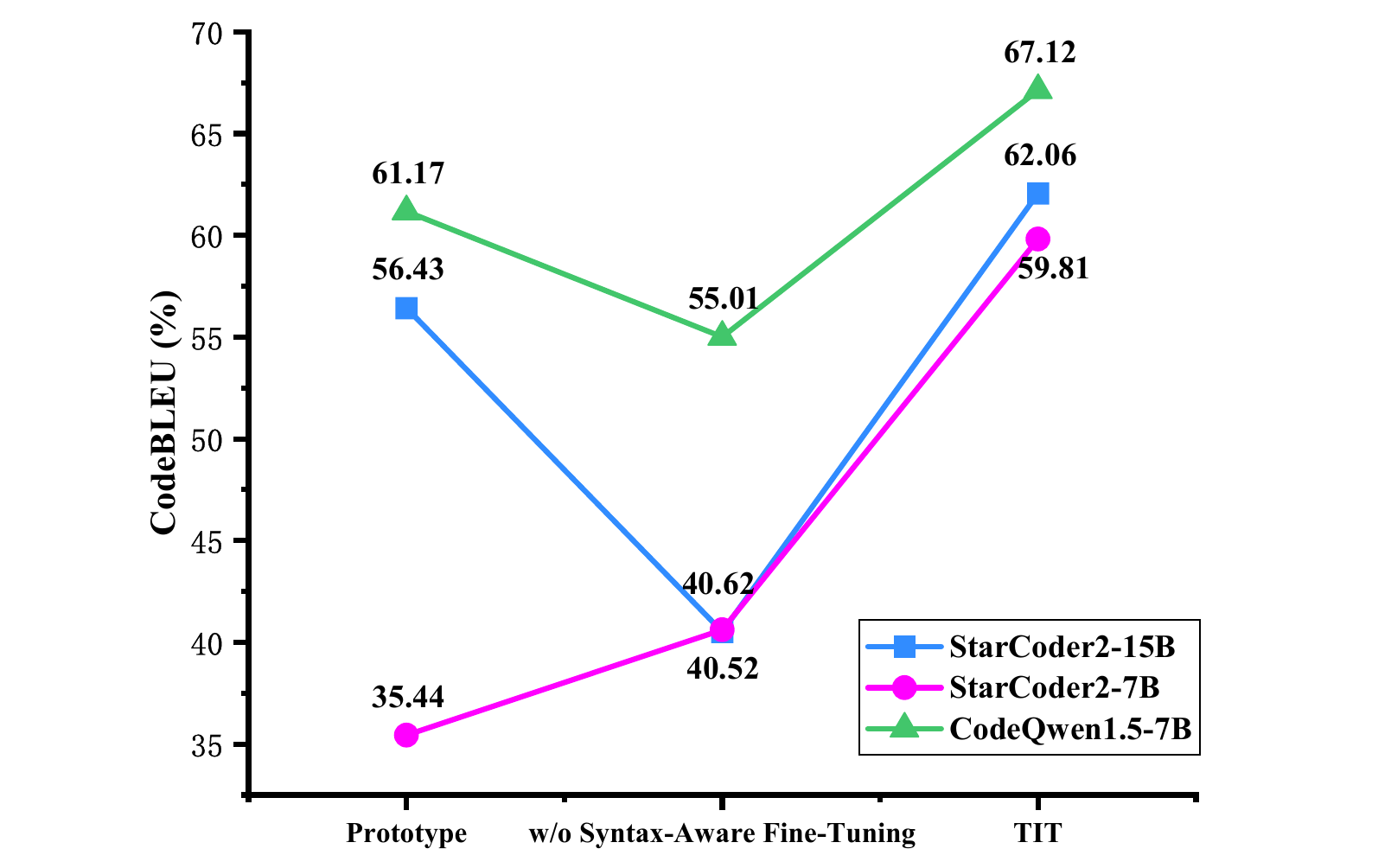}%
\label{fig_sum_5_third_case}}
\caption{The performance of TIT whether using the Syntax-Aware Fine-Tuning. (a) Successful translation rate. (b) Syntactic confusion. (c) CodeBLEU.}
\label{fig_sum_5}
\end{figure}

\textit{3) Dual-Stage Tree Instruction Tuning Module:} To investigate the contribution of the dual-stage tree instruction tuning module to code translation, we compare TIT against an ablated variant without syntax-aware fine-tuning (w/o Syntax-Aware Fine-Tuning). Results are shown in Fig. \ref{fig_sum_5}.

Fig. \ref{fig_sum_5}\subref{fig_sum_5_first_case} shows translation success rates from this ablation study. Results demonstrate that explicit node-to-snippet alignment achieved through TIT's first-stage syntax-aware fine-tuning constitutes the critical mechanism driving translation success. Removing this component causes significant performance drops across all models—for example, StarCoder2-15B drops from 68.29\% to 52.44\% (a 15.86\% absolute decrease). This decline indicates that traditional two-stage frameworks without syntactic grounding only reach 54.93\% success due to overreliance on surface pattern matching. In contrast, TIT's syntax-aware fine-tuning stage binds syntactic information representations to functionality, enabling contextually consistent generation. This approach is especially effective for larger-scale LLMs: StarCoder2-15B shows a 29.72\% absolute improvement over its prototype, compared to 18.95\% for 7B-scale models.

Fig. \ref{fig_sum_5}\subref{fig_sum_5_second_case} illustrates this ablation study's contribution to mitigating syntactic confusion. The results establish that first-stage syntax-aware fine-tuning is essential for high-quality code translation: removing it causes significant regression, with StarCoder2-15B's syntactic confusion increasing by 100\% (from 3 to 6) and a 15.86\% drop in success rate. By explicitly aligning syntactic representations with target code snippets, this tuning helps decrease confusion and supports the generation of contextually consistent output in subsequent translation stages. Conversely, StarCoder2-7B's slight degradation (from 15 to 14, a -7.13\% change) highlights how tuning effectiveness varies with model capacity.

Fig. \ref{fig_sum_5}\subref{fig_sum_5_third_case} shows the CodeBLEU score from this ablation study. Ablation studies under the CodeBLEU metric demonstrate that the first-stage syntax-aware fine-tuning significantly enhances TIT performance. Compared to the variant without this stage, TIT exhibits CodeBLEU score improvements of approximately 53.16\%, 47.25\%, and 22.02\% on StarCoder2-15B, StarCoder2-7B, and CodeQwen-7B, respectively. These results indicate that the first stage strengthens the model's ability to parse unified syntactic information representations, not only enhancing the capture of syntactic dependencies but also improving the consistent reconstruction of source code functional logic and identifier naming. Notably, even large-scale LLMs experience substantial performance degradation when deprived of the first stage, underscoring the architecture-agnostic universal value of syntax-aware fine-tuning in reducing input context redundancy, increasing information density, and generating precise target code.

Collectively, these results demonstrate the critical role of syntax-aware fine-tuning in code translation tasks. This approach enhances the LLMs’ capacity to interpret representations of syntactic structures with specialized patterns while simultaneously alleviating the cognitive load associated with parsing complex syntactic information, ultimately optimizing code translation performance.

This ablation study demonstrates that the fine-grained parallel dataset augmentation module effectively assists LLMs in comprehending syntactic information representations and achieving precise syntax with semantics by providing large-scale, granular training data. Concurrently, the two-stage tree-instruction fine-tuning module systematically guides hierarchical mastery of syntactic structures and code generation capabilities. These core components operate synergistically to enhance model performance in code translation tasks, significantly improving both parsing accuracy for complex syntactic constructs and generation quality of target code.

\section{Threats to Validity}
\subsection{Internal Validity}
Firstly, a potential threat lies in the reproducibility and accuracy of baseline methods. When re-implementing the models, we ensured consistency with the original parameter settings by following the released code. If a baseline did not provide a Python-to-Java translation version, we mitigated this threat by retraining the model on collected parallel data for this task.

Secondly, another threat comes from the construction of fine-grained parallel datasets, where statement-level segmentation and sampling may introduce noise and affect model performance. To address this, we conducted multiple experiments under different sampling ratios and referred to prior findings to control the distribution of statements, showing that our strategy consistently improved translation performance.

Finally, the dual-stage tree instruction tuning introduces additional complexity that may increase risks of overfitting or unstable training. We alleviated this threat by conducting ablation studies on both stages, confirming that the design not only reduces syntactic confusion but also improves translation success rate. Future work will explore alternative scheduling and regularization strategies to further enhance robustness.

\subsection{External Validity}
The primary external validity threat is that our experiments are mainly conducted on Python-to-Java translation tasks, the HumanEval-X benchmark dataset, and a limited set of mainstream LLMs. Therefore, the results may not fully generalize to all programming languages, datasets, or LLMs. To mitigate this threat, we evaluate TIT across LLMs of different parameter scales and architectures, and further demonstrate through mixed experiments of statement-level and function-level data that the effectiveness of TIT arises not only from increased data volume but also from the intrinsic advantages of its design. In future work, we plan to assess TIT on multilingual code translation tasks, real-world development scenarios, and a broader range of model architectures and parameter scales.

\subsection{Construct Validity}
This study primarily adopts translation success rate and CodeBLEU as evaluation metrics. Although these two metrics are widely used in code translation research, they may still fall short of fully capturing all quality dimensions of translation results. To mitigate this limitation, we take the following measures: the translation success rate is used to directly reflect functional correctness, while CodeBLEU is employed to complement it by measuring structural and surface similarity, making the two metrics mutually reinforcing; in RQ3, we conduct fine-grained error categorization and quantitative analysis of syntactic confusion to directly assess the effectiveness of TIT in reducing syntax-related errors; in RQ4, ablation experiments and concrete case comparisons (manual inspection and illustrative examples) are carried out to validate the contribution of each component to translation quality, thereby strengthening our conclusions regarding semantic and functional improvements. For future work, we plan to further broaden the scope of error analysis in code translation results, aiming to achieve a more comprehensive and fine-grained evaluation of translation quality.

\section{Related Work}
This section summarizes relevant research in code translation.
\subsection{End-to-End Code Translation}
Early end-to-end approaches to code translation relied on manually crafted templates integrated with syntactic rules \cite{r3}, \cite{r4}. For example, Nguyen et al. \cite{r5, r6}  incorporated statistical machine translation (SMT) to automatically extract translation rules from statistical relationships in bilingual code corpora. While pioneering, these methods required significant human effort and had inherent limitations in generalization capability.

Recent advancements have shifted toward neural methods, mainly using Transformer-based unsupervised learning \cite{natural_translation}. These techniques gain translation ability by training on large-scale monolingual code datasets, focusing on accurate token prediction. Importantly, their initial designs were not specifically made for code translation. For instance, CodeBERT, an encoder-only model, was first adapted pre-training to code-related tasks through Masked Language Modeling (MLM) and Replaced Token Detection (RTD), achieving promising results on downstream code tasks. Later, Lu et al. \cite{codexglue} expanded CodeBERT to sequence-to-sequence models, allowing basic code translation.

TransCoder achieved a significant milestone as the first dedicated unsupervised code translation model. By combining denoising autoencoding and back-translation, it captured language-specific patterns from monolingual source code for cross-lingual transfer. Building on this, Lachaux et al. \cite{dobf} introduced the DeObfuscation (DOBF) pre-training objective to improve functional understanding, while TransCoder-ST \cite{transcoder-st} utilized automated unit tests to filter invalid translations, constructing fully verified parallel corpora for fine-tuning.

However, these end-to-end methods face two fundamental limitations. First, data scarcity: the lack of large-scale parallel code datasets limits models' capacity to learn accurate syntactic mappings. Second, syntax neglect: over-reliance on token-level patterns diminishes structural awareness, resulting in translations with syntactic errors (e.g., type mismatches or control flow inaccuracies).

\subsection{IR-Based Code Translation}
IR-based methods have emerged to bridge this gap. These methods reduce dependence on parallel datasets by structurally decoupling the translation process, replacing direct source code input with abstract representations. They explicitly encode syntactic constraints within the IR to improve the structural accuracy of translations. Early research \cite{r2} developed rule-based templates that capture class-, method-, and package-level abstractions after source code redesign and preprocessing. By embedding source information into these templates, Sneed et al. \cite{r2} achieved COBOL-to-Java translation by transforming syntactic dependencies into straightforward business logic mappings, achieving modularization of program interfaces.

The development of deep learning has gradually incorporated intermediate representation techniques into automated code translation. TreeBERT \cite{tree-tree} pioneered the use of ASTs to capture syntactic features for automatic code translation, implementing node-to-node tree translation with a Transformer model pre-trained using tree-structured masked language modeling (MLM) and node order prediction. Due to TreeBERT's limited scalability, subsequent research predominantly converts ASTs into linear sequences as auxiliary inputs. For example, Wang et al. \cite{SynCoBERT} processes ASTs through depth-first traversal, concatenating them with code and comments as multimodal inputs to train SynCoBERT, with experiments demonstrating its effectiveness across four downstream tasks, including code translation. Liu et al. \cite{Syntax_and_Domain} improve performance on code migration tasks by incorporating AST structures into graph representations. Huang et al. \cite{code_distillation} capture semantic and structural similarities in ASTs as pivotal representations for cross-language translation, increasing success rates through representation pre-training tasks.

IR-based code translation extends beyond AST or control flow graph representations. Macedo et al. \cite{intertrans} bridges different syntactic and semantic gaps between source and target languages through cross-lingual intermediate translation sequences. In contrast, Tang et al. \cite{explain} employ natural language as an IR, decoupling the translation process into code summarization and generation subtasks. These specialized IR methods mainly rely on high-resource languages as translation pivots, helping to address data scarcity in low-resource language scenarios.

\section{Conclusion}
In this paper, we propose TIT, a tree-structured instruction tuning method designed to address key limitations in LLM-based code translation. TIT integrates the syntactic information representation module, the fine-grained parallel dataset augmentation module, and the dual-stage tree instruction tuning module, effectively mitigating syntactic confusion and fine-grained semantic misalignment issues, thereby significantly improving LLM performance in code translation tasks. Experimental results demonstrate that TIT significantly improves code translation success rates across different LLM models. Specifically, the translation success rates for StarCoder2-15B, StarCoder2-7B, and CodeQwen1.5-7B are 68.9\%, 60.98\%, and 57.32\%, respectively, showing improvements of 1.75×, 1.45×, and 1.22× over their base models. Additionally, TIT substantially mitigates syntactic confusion in all base models, with the syntactic confusion rate decreasing by 72.73\% for StarCoder2-15B, 17.65\% for StarCoder2-7B, and 80.00\% for CodeQwen-7B. This improvement is primarily due to the removal of source-language-specific syntax in the inputs, reducing redundant information that could interfere with the LLM’s inference process. TIT also achieves translation quality comparable to that of larger parameter LLMs even when deployed on relatively smaller models, demonstrating its effectiveness in optimizing mid-to-small scale LLMs. In future work, we plan to observe and optimize other types of errors to further enhance the performance of TIT. Additionally, we plan to explore the performance of TIT across different computing architectures, particularly in distributed and heterogeneous environments, to evaluate its adaptability and effectiveness in complex systems.

\bibliographystyle{unsrt}  
\bibliography{references}  


\end{document}